\documentclass[aps,prd,twocolumn,showpacs,nofootinbib,superscriptaddress,preprintnumbers]{revtex4-1}
\usepackage[utf8]{inputenc}
\usepackage{graphicx,amsmath,amssymb,bbm,subfig,enumitem}
\usepackage[dvipsnames]{xcolor}
\usepackage{lmodern,dsfont,soul}
\usepackage[normalem]{ulem}  
\usepackage{hyperref}
\usepackage{bigstrut}
\hypersetup{colorlinks=true,
	    urlcolor=PineGreen,
	    linkcolor=PineGreen,
	    citecolor=PineGreen}
	    
\newcommand{\eq}{\begin{eqnarray}}
\newcommand{\en}{\end{eqnarray}}

\begin{document}

\title{Three-body spectrum in a finite volume: the role of cubic symmetry}

\author{M.~D\"oring}
\email{doring@gwu.edu}
\affiliation{The George Washington University, Washington, DC 20052, USA}
\affiliation{Thomas Jefferson National Accelerator Facility, Newport News, VA 23606, USA}
\author{H.-W.~Hammer}
\email{Hans-Werner.Hammer@physik.tu-darmstadt.de}
\affiliation{Institut f\"ur Kernphysik, Technische Universit\"at Darmstadt,
64289 Darmstadt, Germany}
\affiliation{ExtreMe Matter Institute EMMI, GSI Helmholtzzentrum f\"ur Schwerionenforschung, 64291 Darmstadt, Germany}
\author{M.~Mai}
\email{maximmai@gwu.edu}
\affiliation{The George Washington University, Washington, DC 20052, USA}
\author{J.-Y.~Pang}
\email{pang@hiskp.uni-bonn.de}
\affiliation{Helmholtz-Institut f\"ur Strahlen- und Kernphysik (Theorie), Universit\"at Bonn, D-53115 Bonn, Germany}
\affiliation{Bethe Center for Theoretical Physics, Universit\"at Bonn, D-53115 Bonn, Germany}
\author{A.~Rusetsky}
\email{rusetsky@hiskp.uni-bonn.de}
\affiliation{Helmholtz-Institut f\"ur Strahlen- und Kernphysik (Theorie), Universit\"at Bonn, D-53115 Bonn, Germany}
\affiliation{Bethe Center for Theoretical Physics, Universit\"at Bonn, D-53115 Bonn, Germany}
\author{J.~Wu}
\email{jiajunwu@hiskp.uni-bonn.de}
\affiliation{Helmholtz-Institut f\"ur Strahlen- und Kernphysik (Theorie), Universit\"at Bonn, D-53115 Bonn, Germany}
\affiliation{Bethe Center for Theoretical Physics, Universit\"at Bonn, D-53115 Bonn, Germany}


\date{\today}
\preprint{JLAB-THY-18-2646}
\begin{abstract}
The three-particle quantization condition is partially diagonalized in the center-of-mass frame by using cubic symmetry on the lattice. To this end, instead of spherical harmonics, the kernel of the Bethe-Salpeter equation for particle-dimer scattering is expanded in the basis functions of different irreducible representations of the octahedral group. Such a projection is of particular importance for the three-body problem in the finite volume due to the occurrence of three-body singularities above breakup. Additionally, we study the numerical solution and properties of such a projected quantization condition in a simple model. It is shown that, for large volumes, these solutions allow for an instructive interpretation of the energy eigenvalues in terms of  bound and scattering states. 
\end{abstract}

\maketitle

\section{Introduction}

Lattice QCD calculations provide an ab-initio access to hadronic processes. 
These calculations are usually performed in a small cubic volume with  periodic 
boundary conditions and require an infinite-volume extrapolation for the comparison to experimental data. Methods to extract infinite-volume bound state energies and two-body phase shifts have been established long 
ago~\cite{Luscher:1986pf, Luscher:1990ux}. In contrast, infinite-volume extrapolations for three-body systems have only been considered recently; yet they are indispensable to understand many systems of high current interest. 
Some examples of such systems are excited exotic and non-exotic states decaying 
into three mesons or excited baryons that are known to have a sizable decay into two pions and a nucleon.

In the last few years, considerable progress has been made in understanding three-body observables in a finite volume and in deriving the three-particle quantization condition -- that is, an equation that determines the finite-volume spectrum of the three-particle system through the parameters of the three-particle $S$-matrix in the infinite 
volume~\cite{pang1, pang2, Mai:2017bge, Briceno:2017tce, Guo:2017crd, Guo:2017ism, Guo:2016fgl, Hansen:2016fzj, Hansen:2015zta, Hansen:2015zga, Meissner:2014dea, Hansen:2014eka, Polejaeva:2012ut, Briceno:2012rv, Roca:2012rx}, see Refs.~\cite{Briceno:2017max, Briceno:2014tqa} for recent reviews. 
Further studies focused on the behavior of three-body systems in a finite volume~\cite{Kreuzer:2008bi, Kreuzer:2009jp, Kreuzer:2010ti, Kreuzer:2012sr,Jansen:2015lha,Aoki:2013cra} and topological effects due to bound subsystems moving in a box~\cite{Bour:2012hn,Bour:2011ef}. Note that, 
as shown in Refs.~\cite{Hansen:2017mnd, Agadjanov:2016mao},
without mapping out the full three-body dynamics explicitly, one can access certain bulk properties of the scattering amplitude (including intermediate three-body channels). 
 For example, the use of an optical potential~\cite{Agadjanov:2016mao} allows 
to access resonances in a finite volume even in the presence of multiple channels and particles.

In Refs.~\cite{pang1, pang2, Mai:2017bge}, simple and transparent procedures 
have been designed, which 
allow to extract the three-body observables in the infinite volume from the measured finite-volume spectrum. The complications arising from coupled two- and three-body channels in a finite volume have also been addressed recently~\cite{Briceno:2017tce}.
Apart from the general formulation, important 
applications have also been considered in the literature. From these, one may single out
the study of the binding energy of a shallow three-body bound state~\cite{Meissner:2014dea,Hansen:2016ync,Meng:2017jgx}, as well as the calculation of the (perturbative) shift of the
three-particle ground state in a finite volume, see, e.g., Refs.~\cite{Beane:2007qr,Sharpe:2017jej}.

The quantization condition is an equation,
whose roots determine the entire finite-volume energy spectrum for a given system. 
Using the spatial symmetries of the problem, it can be diagonalized in
different irreducible representations of the symmetry group. 
This is an analog of the partial-wave
expansion in the infinite volume limit, where the symmetry group is the rotation
group in coordinate space. 
If the underlying interaction is rotationally invariant, the different partial waves 
decouple and the problem simplifies significantly. In this paper, we 
present  a similar expansion on a cubic lattice, which does not exhibit the rotational 
symmetry any more, and apply this expansion
to a three-body model system.

\begin{figure*}[t]
\begin{center}
\includegraphics*[width=1\linewidth]{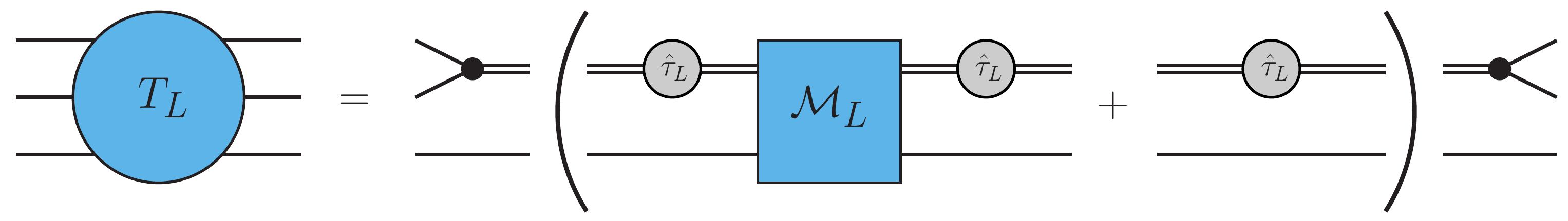}                  
\end{center}
\caption{The three-particle scattering amplitude $T_L$, constructed from
the particle-dimer scattering amplitude ${\cal M}_L$. The double line 
with a blob stands for the dressed dimer propagator. The quantity 
in the parentheses on the right hand side determines the finite volume spectrum 
and will be referred to as $\tilde T_L$ in the main text.\label{fig:T33}}
\end{figure*}

For definiteness, consider the scattering of a two-body bound state with
binding energy $B_2 > 0$ and a third particle at a total center-of-mass
energy $E$ in the infinite volume (we use the usual non-relativistic convention for $E$). 
There are the regions of three-body bound states ($E< -B_2$) and
elastic scattering of the two-body bound state and the third particle
($-B_2 \leq E < 0$). Above the breakup threshold for the two-body bound state ($E=0$), three-body singularities from the free propagation of
three-particle intermediate states appear in the interaction and the corresponding 
scattering amplitude~\cite{SZ74}.
Before the partial-wave projection, these singularities appear as single
poles in the free three-body propagator. They are regulated by the usual
$i\epsilon$ prescription that adds an infinitesimal positive imaginary part to
the energy $E$. 
Integration over the solid angle, required for the 
partial-wave projection, transforms the poles to cuts with logarithmic
branch points in the partial-wave amplitudes. The singularities of the
partial-wave amplitudes are milder than before and the three-body poles in
the original amplitude are restored only when the infinitely many partial
waves are summed up. The branch points are in principle integrable in each
partial wave but they lead to instabilities in numerical calculations
if they are not handled with special care.
In numerical solutions of the three-body scattering problem these cuts are
thus usually treated via contour deformation (if the kernel is known
analytically) or via the introduction of small but finite imaginary
parts in the energy $E$ which are extrapolated to zero in the end.
In practical calculations in the three-nucleon system, the partial-wave expansion then converges well, below and above breakup, see, e.g., Ref.~\cite{Golak:2014ksa}.

In the finite volume matters are different, as the whole spectrum is discrete. In a 
cubic box, only a discrete set of
momenta determined by the boundary conditions (usually taken as periodic)
is allowed. Therefore, the rotational symmetry is broken 
down to cubic symmetry including rotations and reflections, which form the octahedral group $O_h$. The infinitely many irreducible representations of the
rotation group map onto the 10 irreducible representations of the
octahedral group. 
Below the lowest inelastic threshold 
one can still expand in the infinite-volume partial waves, although
they are coupled now due to the breaking of rotational symmetry~\cite{Kreuzer:2008bi,Kreuzer:2009jp,Kreuzer:2010ti,Polejaeva:2012ut,Kreuzer:2012sr,Hansen:2014eka,pang2}. 
The partial-wave expansion for the interaction
converges well in this region, since the regular summation theorem \cite{Luesch86,Doring:2012eu} ensures that the infinite-volume partial-wave projection
remains a good approximation for regular interactions. Above the inelastic thresholds, 
the finite-volume interaction has discrete
poles from three-body intermediate states. 
This averts the convergence of the partial-wave expansion, making the expansion 
in eigenfunctions of the octahedral symmetry group unavoidable in this energy region.
Definition and implementation of such an expansion for the three-body quantization condition is the aim of the current study. We shall in particular demonstrate that, owing to the octahedral symmetry, the three-body quantization condition~\cite{pang1, pang2, Mai:2017bge, Briceno:2017tce, Hansen:2016fzj, Hansen:2015zta, Hansen:2015zga, Hansen:2014eka, Polejaeva:2012ut, Briceno:2012rv, Roca:2012rx}
can be split into different independent equations, whose solutions determine the 
finite-volume energy spectrum in different irreducible representations (irreps) of the octahedral group. This is extremely convenient, since the source/sink operators in lattice calculations are usually chosen to transform as irreducible tensor operators under the octahedral group, allowing one to determine the energy spectra in different irreps independently.

In this paper, we present two equivalent methods to implement the above
expansion with different technical advantages.
The plan of the paper is as follows. In Section~\ref{sec:quantization},
we discuss the three-particle quantization condition and, in particular,
various singularities, appearing there. In Sections~\ref{sec:group} 
and~\ref{sec:alternative},
we describe both diagonalization procedures for the quantization
condition in the basis functions of different irreps. 
Finally, Section~\ref{sec:numerical} contains an example of a numerical 
solution of the quantization condition involving the projection method. 
The interpretation of the energy levels in terms of the bound and scattering states
is also discussed. Technical details are relegated to the appendices.

\section{Three-particle quantization condition}
\label{sec:quantization}

The three-body quantization condition can be derived in
the particle-dimer picture used in Refs.~\cite{pang1,pang2} or the spectator-isobar picture used in Refs.~\cite{Mai:2017bge}. As far as the quantization condition is concerned, the formal differences play no role. Therefore, in the following we refer in many occasions to both as ``dimers'' for simplicity. 
To describe the interactions in the three-particle system within this
picture, one needs the following blocks: the 
two-particle interactions, characterized by the dimers (isobars) and the
interaction of the dimer (isobar) with the third particle (spectator). Two
important remarks are in order. First, the dimer picture is not an {\em approximation,} but an {\em equivalent
description} of the three-particle system, if one allows for dimers of any
spin  and for a generic two-particle-dimer vertex~\cite{pang1,pang2}. 
This corresponds to the partial-wave expansion in the two-body
amplitude.
Second, the dimer formalism does not necessarily imply the presence of
a shallow two-body bound-state (resonance), albeit in this case the use of
this formalism might be more efficient, since one term will dominate the
expansion of the two-body scattering amplitude. Both statements are
self-evident in effective field theory, where the dimer field is just
a dummy integration variable in the path integral~\cite{Bedaque:1998kg,Bedaque:1998km,Bedaque:1999vb}. However, their validity
does not depend, of course, on a particular framework used. For example,
in Ref.~\cite{Mai:2017vot} the two-body interaction in a given partial wave 
is parameterized by an amplitude obtained within the $N/D$ method, with the
analytic properties directly determined from the three-body unitarity. Nevertheless, all final expressions are similar to those obtained in the effective field theory 
framework.

In a finite volume, the energy levels coincide with the position of the
poles of the finite-volume three-particle amplitude. This amplitude can
be related to the particle-dimer scattering amplitude -- schematically, this
relation is depicted in Fig.~\ref{fig:T33}. Since the external vertices
(outside the brackets in Fig.~\ref{fig:T33}) are non-singular, they can be left
out. The remaining piece of the finite-volume amplitude, which is depicted in the parentheses on the right hand side of Fig.~\ref{fig:T33}, reads as 
\begin{align}\label{eq:tildeTL}
\tilde T_L({\bf p},{\bf q};E)=\hat\tau_L({\bf p};E)\, {\cal M}_L( {\bf p},{\bf q};E)
\,&\hat\tau_L({\bf q};E)\\
&+\hat\tau_L({\bf p};E)\,L^3\delta_{{\bf p}{\bf q}}\,,\nonumber
\end{align}
where $L$ denotes the spatial size of the box, $E$ is the total energy of the three 
particles, and ${\bf p}$ and ${\bf q}$ are the discretized relative dimer-spectator 
three-momenta, e.g., ${\bf p}\in\{2\pi{\bf n}/L|{\bf n}\in\mathbb{Z} ^3 \}$.
The particle-dimer scattering amplitude obeys the Bethe-Salpeter equation
\begin{align}\label{eq:ML}
{\cal M}_L({\bf p},{\bf q};E&)=Z({\bf p},{\bf q};E)\\
&+
\frac{1}{L^3}\,\sum_{\bf k} ^\Lambda
Z({\bf p},{\bf k};E)\,\hat\tau_L({\bf k};E)\,
{\cal M}_L({\bf k},{\bf q};E)\,,\nonumber
\end{align}
where $\Lambda$ is the ultraviolet cutoff and 
$\hat\tau_L$ stands for the dressed dimer propagator\footnote{The quantity $\tau_L$ from Refs.~\cite{pang1,pang2} is equal to $\hat\tau_L$ up to normalization: $\hat\tau_L=8\pi\tau_L$. Furthermore, we restrict ourselves to the case of three identical scalar particles with a mass $m$ and
consider only one $S$-wave dimer, albeit the dimer formalism is general enough to include higher 
partial waves~\cite{pang2, Mai:2017bge}.}.
Assuming, for instance, the non-relativistic kinematics as in Refs.~\cite{pang1,pang2}, we have:
\eq\label{eq:tauL}
8\pi\hat\tau_L^{-1}({\bf k};E)&=&k^*\cot\delta(k^*)+S({\bf k},(k^*)^2)\, ,
\\[2mm]
S({\bf k},(k^*)^2)&=&-\frac{4\pi}{L^3}\sum_{\bf l}\frac{1}{{\bf k}^2+{\bf k}{\bf l}+{\bf l}^2-mE}\, ,\nonumber
\en
where $k^*$ is the magnitude of the 
relative momentum of the pair in the rest frame,
\eq
k^*=\sqrt{\frac{3}{4}\,{\bf k}^2-mE}\, .
\en
In Eq.~(\ref{eq:tauL}), unlike Eq.~(\ref{eq:ML}), the momentum sum is implicitly 
regularized by using dimensional regularization and $\delta(k^*)$ is the S-wave 
phase shift in the two-particle subsystem.
The effective range expansion reads
\eq
k^*\cot\delta(k^*)=-\frac{1}{a}+\frac{1}{2}\,r(k^*)^2+O((k^*)^4)\, ,
\label{eq:ERE}
\en
where $a,r$ are the two-body scattering length and the effective
range, respectively.
For simplicity, we shall consider the case, 
when the two-body
interactions are described only by the scattering length $a$. 
The effective range $r$ and higher-order shape parameters are all set
 equal to zero, corresponding to the leading order of the effective field theory for short range-interactions~\cite{Bedaque:1998kg,Bedaque:1998km}.

Finally, the quantity $Z$ denotes the kernel of the Bethe-Salpeter
equation. It contains the one-particle exchange diagram, as well as the local term, corresponding to the particle-dimer interaction (three-particle force).
Again, for simplicity, we shall restrict ourselves to the case of non-derivative
coupling, which is described by a single constant $H_0(\Lambda)$.
The kernel then takes the form
\eq\label{eq:Z}
Z({\bf p},{\bf q};E)=\frac{1}{-mE+{\bf p}^2+{\bf q}^2+{\bf p}{\bf q}}
+\frac{H_0(\Lambda)}{\Lambda^2}\, .
\en
The dependence of $H_0(\Lambda)$ on the cutoff is such that the infinite-volume
scattering amplitude is cutoff-independent. In a finite volume, this ensures
the cutoff-independence of the spectrum.

We would like to stress here that using the effective range approximation,
or restricting ourselves to the non-derivative particle-dimer interactions
does not affect the generality of the following arguments. The equations like
Eq.~(\ref{eq:Z}) are displayed here for the illustrative purpose only.
In Refs.~\cite{Mai:2017vot,Mai:2017bge}, both the two-body scattering amplitude
and the three-body force depend on the momenta in a non-trivial way and the
expression of the kernel gets modified. The main reason is the relativistic formulation of Refs.~\cite{Mai:2017vot,Mai:2017bge} plus the fact that the vertices in general have momentum dependence which would correspond to inclusion of higher orders in the effective field theory language. However, all statements made in the present  section apply in this case as well, since the singularity structure of the kernel remains the same. Specifically, unitarity ensures that the exchange singularity is the \emph{only} one leading to imaginary parts of the interaction. Furthermore, three-body unitarity determines the imaginary parts from which real parts can be obtained using dispersion relations without any regularization issues using sufficiently many subtractions. Of course, the amplitude can be mapped to field theory if desired but there is no need to do so. In that framework, the term $H_0$ in Eq.~(\ref{eq:Z}) plays the role of a term that does not violate unitarity and that can be used to accommodate genuine three-body forces. The dispersive framework of Ref.~\cite{Mai:2017vot} also shows that the isobar picture simply provides a convenient parameterization and can be dropped in favor of 
unitary on-shell $2\to 2$ amplitudes formulated in \emph{any} parameterization, 
plus real-valued three-body forces.

In a finite volume, close to a finite-volume pole of $\mathcal{M}_L$ at $E_i$, the amplitude factorizes and can be written in the form
\eq\label{eq:decom}
\mathcal{M}_{L}({\bf p},{\bf q};E)=\frac{f_{i}({\bf p})f^{*}_{i}({\bf q})}{E-E_{i}}\, .
\en
Substituting this expression into the Eq.~(\ref{eq:ML}) and dropping the
parametric dependence on the momentum ${\bf q}$,
a homogeneous equation can be obtained 
\eq\label{eq:homoeq}
f_{i}({\bf p})=\frac{1}{L^{3}}\sum_{\bf k}^{\Lambda}Z({\bf p},{\bf k};E)\hat{\tau}_{L}({\bf k};E)f_{i}({\bf k}).
\en
Here the inhomogeneous term has been dropped,
because it can never develop a pole in the energy.

The quantization condition is obtained from the above equations in a trivial
fashion. Symbolically, the solution for $\tilde T_L$ is given by:
\eq\label{eq:tildeTL1}
\tilde T_L=(\hat\tau_L^{-1}-Z)^{-1}\, .
\en
The quantity $\tilde T_L$ has poles, when the determinant of the
  inverse of the r.h.s. of
Eq.~(\ref{eq:tildeTL1}) vanishes. This finally gives the quantization
condition we are looking for,
\eq\label{eq:master}
\det(\hat\tau_L^{-1}-Z)=0\, .
\en
The l.h.s. of the above equation defines a function of the total energy $E$,
which, for a fixed $\Lambda$ and $L$,  depends both on the two-body input (the two-body scattering amplitude both above and {\em below} the two-body threshold) 
as well as the three-body input (the non-derivative coupling $H_0(\Lambda)$
plus higher-order couplings). The former input can be independently determined from the simulations in the two-particle sector and extrapolation below threshold. Hence, measuring the 
three-particle energy levels, one will be able to fit the parameters of the three-body force. Finally, using the same equations in the infinite volume with the parameters determined on the lattice, one is able to predict the physical observables in the infinite volume.

An important remark is in order. The finite-volume spectrum of the 
three-particle system is {\em uniquely} determined by the poles of the 
three-particle amplitude $T_L$ (or $\tilde T_L$, because the insertions in 
the external lines are not singular, see Fig.~\ref{fig:T33}). It means that
every root of Eq.~(\ref{eq:master}) corresponds to an energy level, and
{\it vice versa.} On the other hand, as it has been noted independently 
in Refs.~\cite{Mai:2017bge,pang2}, both $\hat\tau_L^{-1}$ and $Z$ become singular 
at certain energies, which leads to the fact that the 
{\em particle-dimer amplitude} ${\cal M}_L$ will contain both, poles 
corresponding to the genuine three-body levels and spurious poles arising
from these singularities. 
In the amplitude $\tilde T_L$ these poles are canceled 
automatically. This was noted and described in detail in 
Ref.~\cite{Mai:2017bge}, see also Refs.~\cite{Polejaeva:2012ut,Briceno:2012rv}.
 This cancellation rests on two essential observations: 
\begin{enumerate}
\item
At the energies, where $Z$ becomes infinite, $\hat{\tau}_L$ is exactly zero. 
Physically, the reason is three-body unitarity, which restricts 
the singularity structure of these quantities~\cite{Mai:2017bge}. 
\item
The dressed propagator of the dimer is singular at real energies, corresponding
to the two-particle energy levels in a finite volume. 
This singularity is canceled in  $\tilde T_L$
if the disconnected piece (the second term on 
the r.h.s. of Fig.~\ref{fig:T33} or of Eq.~(\ref{eq:ML})) is included~\cite{Mai:2017bge}. 
Physically, this is due to the LSZ-reduction formula, which relates the full 
$S$-matrix elements  (including all possible clusterings of transitions) to 
the correlation functions of the underlying field theory
 which, by definition, do not have these singularities. 
\end{enumerate}
The cancellation considered above guarantees that there are no spurious
 poles in the quantization condition, see Eq.~(\ref{eq:master}).

\section{Projection onto the irreps}
\label{sec:group}

As already mentioned, the quantization condition, Eq.~(\ref{eq:master}),
determines the entire spectrum of the three-body system. It would be useful
to partially diagonalize this equation, projecting it to the various irreps
of the octahedral group -- in particular because, in practice,
 the energy levels corresponding to a given irrep are extracted on the 
lattice. Moreover, Eq.~(\ref{eq:master}) may, in general,  contain 
very large matrices. Namely, the matrices $\hat\tau_L$ and $Z$ have dimension
$N ^3\times N ^3$, where $N$ is the largest integer that does not
exceed $\Lambda L/(2\pi)$. The projection of the quantization condition
to a given irrep allows one to reduce the dimension of the matrices.

In this section, we describe how such a projection, which 
qualitatively resembles the partial-wave expansion in the infinite volume, 
can be performed. As discussed in the introduction, 
this becomes necessary due to the occurrence
of three-body singularities.
For two-body systems this step is not required because the two-body
interactions are regular and the regular summation theorem applies.

\subsection{The group and its irreps}

The symmetry group ${\cal G}$ 
of the cubic lattice in the rest frame is the octahedral group that
consists of 24 rotations $R_a$ and the inversion $I$ of all axes. 
These rotations with $a=1,\ldots,24$ 
 can be characterized either by the unit vector 
${\bf n}^{(a)}$ along the rotation axis and angle $\omega_a$, 
or, equivalently, by three Euler angles $\alpha_a,\beta_a,\gamma_a$, 
given, e.g., in table A.1 of Ref.~\cite{Bernard:2008ax}. 
A general element of the group ${\cal G}$ is given by $g=R_aI$, 
where $I$ commutes with all $R_a$, i.e., the total number of the elements of ${\cal G}$ is equal to $48$.
The irreps of the octahedral group of 24 elements (pure rotations) are
\begin{itemize}
\item[$A_1$:]
the trivial one-dimensional representation which assigns $+1$ to all 24 
elements of the group.
\item[$A_2$:]
the one-dimensional representation, which assigns $-1$ to the rotations from the conjugacy classes $6C_4$, $6C_2'$ (see table A.1 of Ref.~\cite{Bernard:2008ax}) and $+1$ otherwise.
\item[$E$:]
the two-dimensional representation. The corresponding matrices are given, e.g., in Eq.~(A.2) of Ref.~\cite{Bernard:2008ax}.
\item[$T_1$:]
the three-dimensional representation. The corresponding matrices are given via the rotation parameters
\begin{align}
T_{\sigma\rho}(R_a)&=\cos\omega_a\delta_{\sigma\rho}\\
&+
(1-\cos\omega_a)n^{(a)}_\sigma n^{(a)}_\rho
-\sin\omega_a\varepsilon_{\sigma\rho\lambda}n^{(a)}_\lambda\, .\nonumber
\end{align}
Here, the indices $\sigma,\rho,\lambda=1,2,3$ and 
$a=1,\ldots,24$ (no inversions). 
\item[$T_2$:]
is the same with a change of sign in the conjugacy classes $6C_4$, $6C_2'$ .
\end{itemize}
For convenience, we collect all these matrices in 
Appendix~\ref{app:group}.

Adding inversions, each of these representation duplicates, e.g., $A_1\to A_1^\pm$, etc. In these representations, the elements corresponding to $R$ and $RI$ are the same for ``$+$'' and have the opposite sign for ``$-$''.

\subsection{Shells}
\label{sec:shells}

Our aim is to carry out an analog of the partial-wave expansion in momentum space in case of cubic symmetry. 
For this reason, we define the {\em shells}, which are the analog of the surface $|{\bf p}|=\mbox{const.}$ in case of the rotational symmetry.
In order to ease notations, we shall measure all momenta in units of $2\pi/L$ in this section.

\begin{itemize}[leftmargin=0.5cm]
\item
The (single) momentum with the smallest length is $(0,0,0)$. This defines the shell
$s=1$. All 48 symmetry transformations, acting on this vector
leave it invariant.
\item
The vectors with the second smallest length are $(1,0,0)$ and all vectors 
that are obtained from this vector by symmetry transformations, which always 
boil down to the  permutation of components of a vector or a
change of sign of one or several components. 
In this case, there are 6 different vectors which define the next shell,
i.e. $s=2$ with the {\em multiplicity} $\vartheta(2)=6$. Furthermore,
 note once more that each vector in a shell is obtained through
\begin{align}
{\bf p}=g\,{\bf p}_0\, ,\quad\quad g\in {\cal G}\, .
\end{align}
We shall refer to ${\bf p}_0$ to as {\em reference vector} of a given shell. 
Nothing depends on the choice of this vector.
\item
One can continue this procedure, consequently including the vectors with a larger length. At some point, one finds that there are vectors with the same length, which are nevertheless not related by the symmetry transformations. First,
 this situation arises for the sets with the reference vectors
$(3,0,0)$ and $(2,2,1)$. We assign such vectors to different
shells, $s=9$ and $s=10$ in this example. 
Thus, our definition of a shell implies that all vectors in a given shell 
are produced from one reference vector by applying symmetry transformations.  
\end{itemize}

At this point, one may consider a spherically symmetric function, which depends
only on the magnitude of ${\bf p}$. The rotation that transforms
$(3,0,0)$ into $(2,2,1)$, albeit not belonging to ${\cal G}$, still leaves this
function invariant -- in other words, the symmetry of this function for a given
$|{\bf p}|$ is larger than ${\cal G}$. It is clear that, going to higher
shells, the number of the combinations of the vectors with the same length 
(in units of $2\pi/L$) is going to increase, rendering the symmetry larger.
One may now fix $|{\bf p}|$ in physical units and increase $L$ -- in a result,
this vector will correspond to the shell(s) with a very high value of $s$ and hence
with a very high degree of degeneracy. The above discussion provides a
qualitative argument in favor of the conclusion that the rotation
symmetry is restored in the limit $L\to\infty$.

\subsection{Expansion in the irreps}
\label{sec:exp1}

An arbitrary function $f({\bf p})$ 
can be characterized by the shell to which the
momentum ${\bf p}$ belongs, and the orientation of ${\bf p}$. Below, we aim at
finding a counterpart of the well-known partial-wave expansion
\begin{align}\label{eq:exp-f}
f({\bf p})&=\sqrt{4\pi}\sum_{\ell m}Y_{\ell m}({\bf\hat p}) f_{\ell m}(p)\,,\\\nonumber
f_{\ell m}(p)&=\frac{1}{\sqrt{4\pi}}\,
\int d\Omega Y^*_{\ell m}({\bf\hat p}) f({\bf p})
\end{align}
for the case of the cubic symmetry on the lattice. In the above expression,
$Y_{\ell m}$ are spherical harmonics, and $p$, ${\bf\hat p}$ denote the magnitude and the unit vector in the direction of the vector ${\bf p}$, respectively.

On the cubic lattice, the analog of the above expansion is given by 
\begin{align}\label{eq:exp}
f({\bf p})=f(g{\bf p}_0)=
\sum_\Gamma\sum_{\rho\sigma} 
T^{\Gamma}_{\sigma\rho}(g)f^{\Gamma}_{\rho\sigma}({\bf p}_0)\,,
\end{align}
where $\Gamma=A_1^\pm,A_2^\pm,E^\pm,T_1^\pm,T_2^\pm$, and the matrices of the irreducible representations 
$T^{\Gamma}_{\sigma\rho}(g)$ are specified in Appendix~\ref{app:group}. 
Note also that
 $f^{\Gamma}_{\rho\sigma}({\bf p}_0)
=f^{\Gamma}_{\rho\sigma}(g{\bf p}_0)$ for
all $g\in {\cal G}$.\footnote{In the analog case of rotational symmetry,
    $f_{\ell m}(p)$ depends only on
    the magnitude $p$ and not the orientation of the momentum ${\bf p}$.}
   
Using the orthogonality of the matrices of the irreducible representations, it is possible to project out the quantity $f^{\Gamma}_{\rho\sigma}({\bf p}_0)$: 
\begin{align}\label{eq:proj}
\sum_{g\in {\cal G}}(T^{\Gamma}_{\lambda\delta}(g))^*f(g{\bf p}_0)&=
\sum_{g\in {\cal G}}(T^{\Gamma}_{\lambda\delta}(g))^*
\sum_{\Gamma'}\sum_{\rho\sigma}
 T^{\Gamma'}_{\sigma\rho}(g)f^{\Gamma'}_{\rho\sigma}({\bf p}_0)
\nonumber\\[2mm]
&=\sum_{\Gamma'}\sum_{\rho\sigma}\frac{G}{s_\Gamma}\,\delta_{\Gamma\Gamma'}
\delta_{\sigma\lambda}\delta_{\rho\delta}f^{\Gamma'}_{\rho\sigma}({\bf p}_0)
\nonumber\\[2mm]
&=\frac{G}{s_\Gamma}\,f^{\Gamma}_{\delta\lambda}({\bf p}_0)\, ,
\end{align}
where $G=48$ is the total number of elements in the group ${\cal G}$ and $s_\Gamma$
is the dimension of the representation $\Gamma$:
$s_\Gamma=1$ for $\Gamma=A_1^\pm,A_2^\pm$, $s_\Gamma=2$ for $\Gamma=E^\pm$,
$s_\Gamma=3$ for $\Gamma=T_1^\pm,T_2^\pm$.

In order to clarify the meaning of the above expressions, let us assume that
$f({\bf p})$ is a regular function which can be expanded in partial
waves, see Eq.~(\ref{eq:exp-f}). Substituting this expansion into 
Eq.~(\ref{eq:proj}), one obtains
\begin{align}\label{eq:proji}
\frac{G}{s_\Gamma}f^{\Gamma}_{\rho\sigma}({\bf p}_0)
=\sqrt{4\pi}\sum_{\ell m}\sum_{g\in {\cal G}}(T^{\Gamma}_{\sigma\rho}(g))^*
Y_{\ell m}(g{\bf\hat p}_0)f_{\ell m}(p)\,.
\end{align}
Here, we used the fact that $|g{\bf p}_0|=|{\bf p}_0|=p$. 
Furthermore, we use the relation
\eq
Y_{\ell m}(g{\bf\hat p}_0)=\sum_{m'}\hat D^{(\ell)}_{mm'}(g)Y_{\ell m'}({\bf\hat p}_0)\, ,
\en
where the quantity $\hat D^{(\ell)}_{mm'}(g)$ is defined as follows:
if $g$ is a pure rotation, then  $\hat D^{(\ell)}_{mm'}(g)$
coincides with the conventional Wigner matrix $D^{(\ell)}_{mm'}(g)$;
if $g$ contains the inversion, it can be represented as $g=R_aI$.
Then, $\hat D^{(\ell)}_{mm'}(g)=(-1)^\ell D^{(\ell)}_{mm'}(R_a)$.

It can be seen now that the vectors
\eq\label{eq:xi}
\xi_{\rho\sigma m}^{\ell\Gamma}({\bf\hat p}_0)
=\sum_{g\in {\cal G}}(T^{\Gamma}_{\sigma\rho}(g))^*
\sum_{m'}\hat D^{(\ell)}_{mm'}(g)Y_{\ell m'}({\bf\hat p}_0)
\en
coincide with the basis vectors of the 
irrep $\Gamma$, corresponding to a given angular
momentum $\ell$, see e.g., Refs.~\cite{Bernard:2008ax,Gockeler:2012yj}.
Namely, the indices $\sigma,\rho$ label the basis vectors, and
$m$ the components of each vector. Consequently, the quantity
$f^{\Gamma}_{\rho\sigma}({\bf p}_0)$ can be expanded in the basis vectors of 
a particular irrep $\Gamma$:
\eq
f^{\Gamma}_{\rho\sigma}({\bf p}_0)=\frac{{\sqrt{ 4\pi}}s_\Gamma}{G}\,
\sum_{\ell m}f_{\ell m}(p)\xi_{\rho\sigma m}^{\ell\Gamma}({\bf\hat p}_0)\, .
\en
In Section~\ref{sec:alternative}, we will describe an alternative
      method to construct a minimal basis for a given shell $s$.

\subsection{The projection of the quantization condition}
\label{sec:reduction}

The quantization condition can be obtained starting from~(\ref{eq:homoeq}).
Hiding the dependence on the energy $E$ and level index $i$, we obtain
\begin{align}
\label{eq:LS3}
f({\bf p})=\frac{1}{L^3}\,\sum_{\bf k} Z({\bf p},{\bf k})\hat\tau_L({\bf k})f({\bf k})\,.
\end{align}
Moreover, the quantities $Z$ and 
$\hat\tau_L$
are scalars with respect to the group ${\cal G}$. This means that for all $g\in {\cal G}$
\begin{align}
Z(g{\bf p},g{\bf k})=Z({\bf p},{\bf k})\,\quad \text{and} \quad
\hat\tau_L(g{\bf k})=\hat\tau_L({\bf k})\,.
\end{align}
The summation over the lattice momenta ${\bf k}$ can be replaced by the summation over the shells $s$ and over the orientations of the momentum ${\bf k}$ inside a given shell $s$, which are described by ${\bf k}(s)=g{\bf k}_0(s),~g\in {\cal G}$ (here, we explicitly indicate the shell index $s$). Altogether, we have $\vartheta(s)$ different vectors, but $G$ terms in the
sum over all elements of the group ${\cal G}$, so each vector will appear $G/\vartheta(s)$ times in this sum. Taking this fact into account, we may rewrite Eq.~(\ref{eq:LS3}) in the following form
\begin{align}\label{eq:LSg}
f({\bf p})=\frac{1}{L^3}\,\sum_s\sum_{g\in {\cal G}}\frac{\vartheta(s)\hat\tau_L(s)}{G}\,
 Z({\bf p},g{\bf k}_0(s))f({\bf k})\, .
\end{align}
Here, we used the fact that $\hat\tau_L({\bf k})$ does not depend on the orientation of ${\bf k}$ in a given shell, i.e., that it is a function of the shell index only.

Multiplying now this equation with $(T^{\Gamma}_{\sigma\lambda}(g'))^*$ from the left
and using Eqs.~(\ref{eq:exp}, \ref{eq:proj}), we obtain
\begin{align}\label{eq:LSinter}
\frac{G}{s_\Gamma}\,f^{\Gamma}_{\lambda\sigma}(r)
=\frac{1}{L^3}\,\sum_s\frac{\vartheta(s)\hat\tau_L(s)}{G}\, \sum_{\Gamma'}\sum_{\rho\delta}
Z^{(\Gamma\Gamma')}_{\lambda\sigma,\rho\delta}(r,s)f^{\Gamma'}_{\rho\delta}(s)\, ,
\end{align}
where we have adjusted a notation
$f^{\Gamma}_{\lambda\sigma}({\bf p}_0(r))\to f^{\Gamma}_{\lambda\sigma}(r)$. Furthermore,
 in the above equation,
\begin{align}\label{eq:longproj}
&Z^{(\Gamma\Gamma')}_{\lambda\sigma,\rho\delta}(r,s)
=\sum_{g,g'\in {\cal G}}(T^{\Gamma}_{\sigma\lambda}(g'))^*Z(g'{\bf p}_0(r),g{\bf k}_0(s))
T^{\Gamma'}_{\delta\rho}(g)
\nonumber\\[2mm]
&=\sum_{g,g'\in {\cal G}}(T^{\Gamma}_{\sigma\lambda}(g'))^*Z(\underbrace{g^{-1}g'}_{=g''}{\bf p}_0(r),{\bf k}_0(s))
T^{\Gamma'}_{\delta\rho}(g)
\nonumber\\[2mm]
&=\sum_{g,g''\in {\cal G}}(T^{\Gamma}_{\sigma\lambda}(gg''))^*Z(g''{\bf p}_0(r),{\bf k}_0(s))
T^{\Gamma'}_{\delta\rho}(g)
\nonumber\\[2mm]
&=\sum_{g,g''\in {\cal G}}\sum_\omega(T^{\Gamma}_{\sigma\omega}(g))^*(T^{\Gamma}_{\omega\lambda}(g''))^*Z(g''{\bf p}_0(r),{\bf k}_0(s))
T^{\Gamma'}_{\delta\rho}(g)
\nonumber\\[2mm]
&=\sum_{g''\in {\cal G}}\sum_\omega\frac{G}{s_\Gamma}\,\delta_{\Gamma\Gamma'}
\delta_{\sigma\delta}\delta_{\omega\rho}
(T^{\Gamma}_{\omega\lambda}(g''))^*Z(g''{\bf p}_0(r),{\bf k}_0(s))
\nonumber\\[2mm]
&=\frac{G}{s_\Gamma}\,\delta_{\Gamma\Gamma'}\delta_{\sigma\delta}
\sum_{g\in {\cal G}}(T^{\Gamma}_{\rho\lambda}(g))^*Z(g{\bf p}_0(r),{\bf k}_0(s))
\nonumber\\[2mm]
&\doteq\frac{G}{s_\Gamma}\,\delta_{\Gamma\Gamma'}\delta_{\sigma\delta}
Z^{\Gamma}_{\lambda\rho}(r,s)\, .
\end{align}
Using this result, we can rewrite Eq.~(\ref{eq:LSinter}) as
\begin{align}
f^{\Gamma}_{\lambda\sigma}(r)
=\frac{1}{L^3}\,\sum_s\frac{\vartheta(s)\hat\tau_L(s)}{G}\, \sum_\rho
Z^{\Gamma}_{\lambda\rho}(r,s)f^{\Gamma}_{\rho\sigma}(s)\, .
\end{align}
Note that, due to the symmetry, 
$Z^{\Gamma}_{\lambda\rho}(r,s)$ and hence $f^{\Gamma}_{\rho\sigma}(s)$ do not depend on $\sigma$.

Finally, the quantization condition in a given irrep $\Gamma$ takes the form
\begin{align}
\det\biggl(\hat\tau_L(s)^ {-1}\delta_{rs}\delta_{\sigma\rho}
-\frac{\vartheta(s)}{GL^3}\,
Z^{\Gamma}_{\sigma\rho}(r,s)\biggr)=0\, .\label{eq:qc-gnl}
\end{align}
Note that, on a given shell, the quantity $Z^{\Gamma}_{\sigma\rho}(r,s)$ may vanish for certain $\Gamma$. A trivial example: as seen from Eq.~(\ref{eq:longproj}), all sums 
except $\Gamma=A_1^+$ vanish on the first shell $r=1$ or $s=1$. In the calculation
of the determinant, one could first ``compress'' the matrix $Z^{\Gamma}_{\sigma\rho}(r,s)$ by deleting all rows/columns which consist only of zeros.
The three-particle quantization 
condition, projected onto the different irreps, which is displayed 
in Eq.~(\ref{eq:qc-gnl}), represents our main result.
The shells are truncated by using a sharp cutoff at
  $N=\Lambda L/(2\pi)$. We emphasize that the solution of the quantization condition,
  Eq.~(\ref{eq:qc-gnl}), is cutoff-independent.
  This happens because the cutoff-dependence of the effective couplings
  $H_{0}(\Lambda),H_{2}(\Lambda),\cdots$ ensures that the physical observables
  in the infinite volume are cutoff-independent. At the same time,
  the finite-volume spectrum becomes cutoff-independent as well, since at short
  distances (of order of $1/\Lambda$), the effect of a finite-size box is not felt.

In a concluding remark we address the issue of partial-wave mixing in  
Eq.~(\ref{eq:qc-gnl}). There are two
angular momenta in the problem: the angular momentum $\ell$ of the  
spectator-dimer system
and the internal angular momentum $\ell'$ of the dimer (the dimer  
spin). The dimer spin $\ell'$
corresponds to the angular momentum of the interacting particle pair  
and is kept as zero
throughout this paper, while $\ell$ can be arbitrary. Obviously, the  
$\ell'=0$ particle-particle
interactions generate contributions corresponding to all values of  
$\ell$ in the
projected interaction kernel $Z^\Gamma$ in Eq.~(\ref{eq:qc-gnl}).
Therefore, if one truncates the expansion of the polynomial term at $\ell=0$ as in Eq. (\ref{eq:Z}),  
the energy levels in Eq.~(\ref{eq:qc-gnl}) are determined by the S-wave couplings only.

The inclusion of the higher partial waves proceeds in the standard fashion described, e.g., in Refs.~\cite{pang1,pang2}. On the opposite, if one expands in the conventional spherical functions instead of the basis functions of the irreps of the octahedral group, as was done, e.g., in 
Refs.~\cite{Kreuzer:2008bi,Kreuzer:2009jp,Kreuzer:2010ti,Kreuzer:2012sr,pang2},
one gets the partial-wave mixing already in the presence of the S-wave 
couplings only. It is clear that this mixing is hand-made and can be avoided,
if the method described in this section is used.


\section{Expansion in cubic harmonics}
\label{sec:alternative}

In the following, we describe an alternative method for the 
projection of the scattering amplitudes to different irreps which resembles the usual projection to partial waves in the infinite volume. 
Such a procedure was already applied  
in Ref.~\cite{Mai:2017bge} for the irrep $A_1^+$, but is generalized here. 
Finally, we will make 
a comparison of this method with the one presented in the previous section.

The momentum shells in the new approach are defined as in Section~\ref{sec:shells}. Furthermore, we define the finite-volume {\em scalar product} for any two functions $f$ and $g$ on a given shell $s$ by
\begin{align}
\langle f,g\rangle_s=\frac{4\pi}{\vartheta(s)}\, \sum_{j}^{\vartheta(s)}
f({\bf\hat p}_j)^*g({\bf\hat p}_j)\, ,
\end{align} 
where the sum runs over the $\vartheta(s)$ different orientations of the unit vector ${\bf\hat p}_j$ pointing to point $j$ in a given shell.

Our aim is to construct an orthonormal basis with respect to this scalar product, 
which allows one to expand an arbitrary function $f({{\bf\hat p}})$ into a linear 
combination of the basis vectors on a given shell. Each vector is a function of the 
${\bf\hat p}_j$ (we therefore refer also to ``basis functions''). In fact, this is an analog of the standard partial-wave expansion, 
Eq.~(\ref{eq:exp-f}), with some  significant differences. First of all, in contrast to the basis functions on the unit sphere in infinite volume (such as spherical harmonics $Y_{\ell m}({\bf\hat p})$), the full set of basis functions in finite volume is given by a {\em finite} set of the so-called {\em cubic harmonics} 
$X_\ell^{\Gamma\nu \alpha}({\bf\hat p})$ where $\Gamma$  denotes, as before, the irrep 
of the octahedral group and $\alpha$ specifies the  basis vector in the 
given irrep. The additional indices $\ell$ and $\nu$ 
specify 
the angular momentum (see Eq.~(\ref{eq:cubic-hamonics})) and the degeneracy 
at that $\ell$, respectively.

The cubic harmonics are linear combinations of spherical harmonics,
see, e.g., Refs.~\cite{Bernard:2008ax,Gockeler:2012yj} and can be obtained by using
the projection operators defined in Eq.~(\ref{eq:xi}) -- more precisely,
they can be identified with different components (labeled by the index $m$)
of the vector $\xi$ in Eq.~(\ref{eq:xi}). The final result reads
\begin{align}
X_\ell^{\Gamma\nu \alpha}({\bf\hat p})=
\sum_m c_{\ell m}^{\Gamma\nu \alpha}Y_{\ell m}({\bf\hat p})\,.
\label{eq:cubic-hamonics}
\end{align}
These functions are orthogonal in $\Gamma$ and $\alpha$ with respect to the 
infinite-volume scalar product,
 and the
$c_{\ell m}^{\Gamma\nu \alpha}$ are Clebsch-Gordan coefficients.

Evidently, on any shell the number of points cannot be larger than the number of 
elements in the cubic group $G=48$, and it changes from shell to shell according to 
Sec.~\ref{sec:shells}. A useful observation in this respect is that all shells can be characterized by the following seven types
\begin{align}
(000),~(00b),~(0bb),~(bbb),~(0bc),~(bbc),~(bcd) \ , 
\end{align}
where all non-negative integers $0$, $b$, $c$, and $d$ are different, $x\neq y$ for $x,\,y\in\{0,b,c,d\}$.  Each type specifies one point of the shell which through the symmetry transformations of the cubic group gives rise to all points in the shell. The multiplicity $\vartheta$ is unique for all shells of a given type as specified in Tab.~\ref{tab:shell}.

Let us now start constructing the
basis functions for each individual shell. Clearly, any function defined on 
the shell of type $(000)$ does not depend on momenta at all and is proportional 
to the cubic harmonic  $X_0^{A_1^+11}$ which, therefore, is the sole basis
vector on this shell.
Other types of shells contain more points (are of higher multiplicity $\vartheta$), which
coincides with the size of the maximal linear independent 
set of cubic harmonics. 
It is possible to make a unitary transformation such that all cubic harmonics are manifestly real.
Conveniently one can use the following iterative procedure to determine such a set. We begin with the set $S:=\{X_ 0^{A_ 1^+ 11}\}$ and
 define the matrix
\begin{align}
r_{mn}^s=\langle  X_m, X_n\rangle_s
~~\text{for}~~
 X_m,  X_n\in S\,,
\end{align}
where the matrix $r$ depends explicitly on the shell index $s$. The indices 
$m$ and $n$ are cumulative, i.e., they comprise 
$\Gamma$, $\nu$, $\alpha$, and $\ell$. 
The set $S$ contains only one element, therefore the rank of the above $1\times 1$
matrix is $1$.  
If the multiplicity of the shell is also $1$ (for type $(000)$), 
then the set is 
complete and one stops here. If not, we add successively other cubic harmonics
with increasing $\ell$ and different
$\Gamma$, $\alpha$ and $\nu$ to set $S$, each time only keeping those,
 which increase the rank of the matrix $r$, and omitting the rest. 
When the rank of the matrix $r$ reaches the multiplicity of the 
given shell, the set $S$ contains the maximal number of linearly independent 
vectors and the procedure terminates -- adding new 
harmonics cannot increase the rank anymore\footnote{Note that because $X_\ell^{\Gamma\nu i}({\bf\hat p})$ is not diagonal in $\ell$ w.r.t the above scalar product, such a set is not unique. Specifically, one can use the same procedure, but starting from a different $\ell>0$.}. For the afore-defined procedure, the cubic harmonics of $\ell\leq 9$ are required to build maximal sets on each of the seven types of shells.
This is seen in  Tab.~\ref{tab:shell}, which is given in Appendix~\ref{app:cubic}.

It is important to note also 
that, according to the Wigner-Eckart theorem, the matrix
$r^s$ is diagonal in the indices $\Gamma$ and $\alpha$, taking the block-diagonal form:
\eq
r^s_{mn}=\delta_{\Gamma\Gamma'}\delta_{\alpha\alpha'}r^{\Gamma s}_{uu'}\, ,
\en
where the matrix $r^{\Gamma s}_{uu'}$ does not depend on $\alpha$, 
and $u{}^{(}{'}{}^{)}$ denotes a generalized index collecting combinations of $\nu{}^{(}{'}{}^{)}$ and $l{}^{(}{'}{}^{)}$. Thus, the above-described procedure can be carried out for the each irrep $\Gamma$ separately -- the different irreps do not talk to each other.

Finally, the orthonormal basis 
\begin{align}
U_s:=\{\chi_m({\bf\hat p})|m=1,..,\vartheta(s)\}
\label{eq:Bs}
\end{align}
on a given shell $s$ can be constructed by orthonormalizing the linearly 
independent $X_m$ with the use of the matrix $\zeta^s:=(r^s)^{-1/2}$ as
\begin{align}
\chi_m({\bf\hat p})\doteq
\chi^{\Gamma\alpha s}_u({\bf\hat p})
=\sum_n(\zeta_{mn}^s)^* \hat X_n({\bf\hat p})\,,
\end{align}
where $m,n=1,..,\vartheta(s)$, while $u$ labels now the basis vectors for a given $\Gamma$ and $\alpha$ on shell $s$ and
\begin{align}
\langle\chi_m,\chi_n\rangle_s
=\sum_{m',n'}\zeta^s_{mm'}\,r^s_{m'n'}\,(\zeta^s)^{*}_{nn'} 
=\delta_{mn}\,.
\end{align}

Finally, we note that, while the orthonormalization procedure depends explicitly 
on the shell index $s$, the maximal set of linear independent cubic harmonics, 
see Tab.~\ref{tab:shell}, depends only on the type of the shell. As an example, 
Tab.~\ref{tab:shell} in Appendix~\ref{app:cubic}
shows that, for the shell type $(bcd)$,
$\Gamma=T_1^+$, and $\alpha=1$, there are three cubic harmonics participating; 
their linear superpositions lead to the three basis vectors 
$\chi_u^{T_1^+ 1 s}$ with $u=1,\dots,3$. In general, examining this table,
one may conclude that the {\em maximum} number of the linearly independent basis
vectors $\chi_u^{\Gamma \alpha s}$ for a given $\Gamma$ and $\alpha$ for all types 
of shells is given by the dimension $s_\Gamma$ of the irrep $\Gamma$. The basis 
vectors for the first 200 shells are numerically provided in the supplemental material of 
this manuscript. 

With the orthonormal basis $U_s$ on a given shell $s$, any function defined on the points of this shell can be expanded similarly to the partial-wave projection Eq.~\eqref{eq:exp-f} in infinite volume. We indicate the function defined at momentum ${\bf p}_j=|{\bf p}|{\bf\hat p}_j$ as 
$f^s({\bf\hat p}_j)$ where ${\bf p}$ is the momentum associated with shell $s$. The expansion in basis functions reads now
\begin{align}\label{eq:exp-f-finvol_NEW}
f^s({\bf\hat p}_j)&=\sqrt{4\pi}\sum_{\Gamma\alpha}\sum_u
f^{\Gamma \alpha s}_u \chi_u^{\Gamma \alpha s}({\bf\hat p}_j)\,,
\\[2mm]\nonumber
f^{\Gamma \alpha s}_u&=\frac{\sqrt{4\pi}}{\vartheta(s)}\,\sum_{j=1}^{\vartheta(s)}  f^s({\bf\hat p}_j)\chi^{\Gamma \alpha s}_u({\bf\hat p}_j)\,
~~\text{for}~~ 
\chi^{\Gamma \alpha s}_u({\bf\hat p})\in U_s\,,
\end{align}
where $U_s$ is defined in Eq.~(\ref{eq:Bs}) and the sum over $u$ is restricted to the number of basis vectors for a given $\Gamma$, $\alpha$, $s$.

The dimer-spectator amplitude (see Eq.~\eqref{eq:tildeTL} as well as Ref.~\cite{Mai:2017bge}) depends on both incoming and outgoing momenta ${\bf p}$ and ${\bf p}'$, therefore a decomposition into irreps for both  momenta is required. We consider here this 
obvious generalization, assuming rotational symmetry for the transition $Z^{ss'}({\bf\hat p}_j,{\bf\hat p}_{j'})$ as done throughout this paper.
Here, ${\bf\hat p}_{j}$ (${\bf\hat p}_{j'}$) is the direction of point $j$ ($j'$) on shell $s$ ($s'$).
Due to rotational invariance and Schur's lemma,
 the projections to the two- and three-dimensional irreps do 
not depend on the respective basis vectors $\alpha$. The result can then be written as
\begin{align}\label{eq:exp-f-finvol_both_sides}
&Z^{ss'}({\bf\hat p}_j,{\bf\hat p}_{j'})
=4\pi\sum_{\Gamma\alpha}\sum_{uu'}
\chi_{u}^{\Gamma \alpha s}({\bf\hat p}_j)
Z^{\Gamma ss'}_{uu'} \chi_{u'}^{\Gamma \alpha s'}({\bf\hat p}_{j'})\,,
\\\nonumber 
&Z^{\Gamma ss'}_{uu'}
=\frac{4\pi}{\vartheta(s)\vartheta(s')}\,\sum_{j=1}^{\vartheta(s)}
 \sum_{j'=1}^{\vartheta(s')} \chi^{\Gamma \alpha s}_{u}({\bf\hat p}_j)
Z^{ss'}({\bf\hat p}_j,{\bf\hat p}_{j'})
\chi^{\Gamma \alpha s'}_{u'}({\bf\hat p}_{j'})\,
\end{align}
for $\chi^{\Gamma \alpha s}_u({\bf\hat p})\in U_s$. Moreover,
the dependence on the total scattering energy is not explicitly
displayed and $\alpha$ is arbitrary.
Note that $Z^{\Gamma ss'}_{uu'}$ 
still depends on the indices $u$ and $u'$, 
corresponding to the incoming and outgoing momenta. In general, 
one has a (trivial) coupled-channel problem in these indices, similar to Eq.~(\ref{eq:qc-gnl}). 
As we have seen, the maximum number of the coupled equations on a given shell and in a given irrep $\Gamma$ is given by $s_\Gamma$, like in Eq.~(\ref{eq:qc-gnl}).
As mentioned before, the supplemental material to this manuscript provides all needed input (basis vectors, direction of points, correspondence between $s$ and $|{\bf p}|$) to make the numerical implementation of Eq.~(\ref{eq:exp-f-finvol_both_sides}) very convenient.

For completeness, we also quote the general projection of  the quantization condition for the three-body method of Ref.~\cite{Mai:2017bge} (there, only the projection to $A_1^+$ was considered). With Eq.~(\ref{eq:exp-f-finvol_both_sides}) and Eq.~(17) of Ref.~\cite{Mai:2017bge} one obtains the projection of the quantization condition onto the different irreps, adapted to the current notation, 
\begin{align}
\label{eq:qcond2}
\det\left(B^{\Gamma ss'}_{uu'}(W^2) 
+
\frac{2E_s \,L^3}{\vartheta(s)}
\tau_s(W^2)^{-1}\delta_{ss'}\delta_{uu'}\right)=0\, ,
\end{align}
where the relativistic energy is given by 
$E_s=\sqrt{m^2+{\bf p}^2}$, with $|{\bf p}|$ being the momentum associated with shell $s$, and $W$ is the total energy of the three-particle system. The dimer-spectator interaction kernel $B^{ss'}({\bf\hat p}_j,{\bf\hat p}_{j'};W^2)$ and dimer propagator $\tau_s(W^2)$ are given by Eq.~(3) and Eq.~(12) of Ref.~\cite{Mai:2017bge}, respectively. The rows (columns) of the matrices in Eq.~(\ref{eq:qcond2}) are labeled by the indices $s$ and $u$ ($s'$ and $u'$).

The quantization conditions Eq.~\eqref{eq:qc-gnl} derived in 
  Sec.~\ref{sec:group} and Eq.~\eqref{eq:qcond2} above are equivalent modulo the use of relativistic kinematics and higher-order terms in the kernel, see the discussion after Eq.~(\ref{eq:Z}).
Comparing the two approaches, 
we note that the decomposition defined by Eq.~(\ref{eq:exp}) is the same for all 
shells. The price to pay for the generality of
the method presented in Sec.~\ref{sec:group} is that the proposed basis may be too large for a given shell. This does not preclude one from carrying out the reduction of the quantization condition, but may lead to the situation that many entries in the matrix $Z^{\Gamma}_{ij}(r,s)$ are equal to zero (see the discussion at the end of Sec.~\ref{sec:reduction}). The approach of the present section relies on the maximal linearly independent set of cubic harmonics, which are determined universally for all shells, ordered by the 7 types described before (see Table~\ref{tab:shell}). However, the orthonormalization of  such a set (determination of the basis $U_s$) has to be carried out on each set separately. In summary, on a given shell, the basis vectors,  selected from the abundant set
of all vectors in the first approach (\ref{sec:reduction}),
are related to the basis vectors $U_s$ by a unitary transformation.

\section{Numerical calculations}
\label{sec:numerical}

\subsection{Description of the model}

In the previous sections, we have projected the three-body quantization 
condition onto the different irreps of the octahedral group, see Eq.~(\ref{eq:qc-gnl}). 
In this section, we wish to 
demonstrate, how this equation can be solved, and to discuss the properties
of the finite-volume spectrum both below and {\em above} the breakup threshold.
To this end, we perform the calculations in the toy model, already
used for the same purpose in Refs.~\cite{Kreuzer:2008bi,Kreuzer:2009jp,Kreuzer:2010ti,Kreuzer:2012sr,pang2}. This model
corresponds to the leading order of the effective field theory for short range-interactions~\cite{Bedaque:1998kg,Bedaque:1998km}.
Moreover, it gives us an opportunity to compare our
results with previous calculations.  

The toy model, used here, has two coupling constants. First, the two-body
scattering phase shift is given solely in terms of the scattering length
$a$ through $k\cot\delta(k)=-1/a$. Second, the particle-dimer interaction
contains no derivatives and is described by a single coupling constant
$H_0(\Lambda)$ ($\Lambda$ stands for the ultraviolet cutoff). 
The dimer propagator $\hat\tau_L$ and the kernel $Z$ in this
model are given by Eqs.~(\ref{eq:tauL}) and (\ref{eq:Z}), respectively.
What remains is to fix the values of all free parameters and carry out
the calculation of the energy spectrum.

As in Refs.~\cite{Kreuzer:2008bi,Kreuzer:2009jp,Kreuzer:2010ti,Kreuzer:2012sr,pang2}, we assume $m=1$ and $a=1$. All length scales are given in 
units
of $a$. We further fix the cutoff $\Lambda=225$, assuming that it is
high enough for all effects
of order of $1/\Lambda$ to
be neglected. One may now fix the dimensionless 
coupling $H_0(\Lambda)$ e.g., by demanding that the three-body bound state
has a given binding energy $B_3$. Assuming $B_3=10$ gives $H_0(\Lambda)=0.192$.

\begin{figure}[t]
\begin{centering}
\includegraphics[width=1\linewidth]{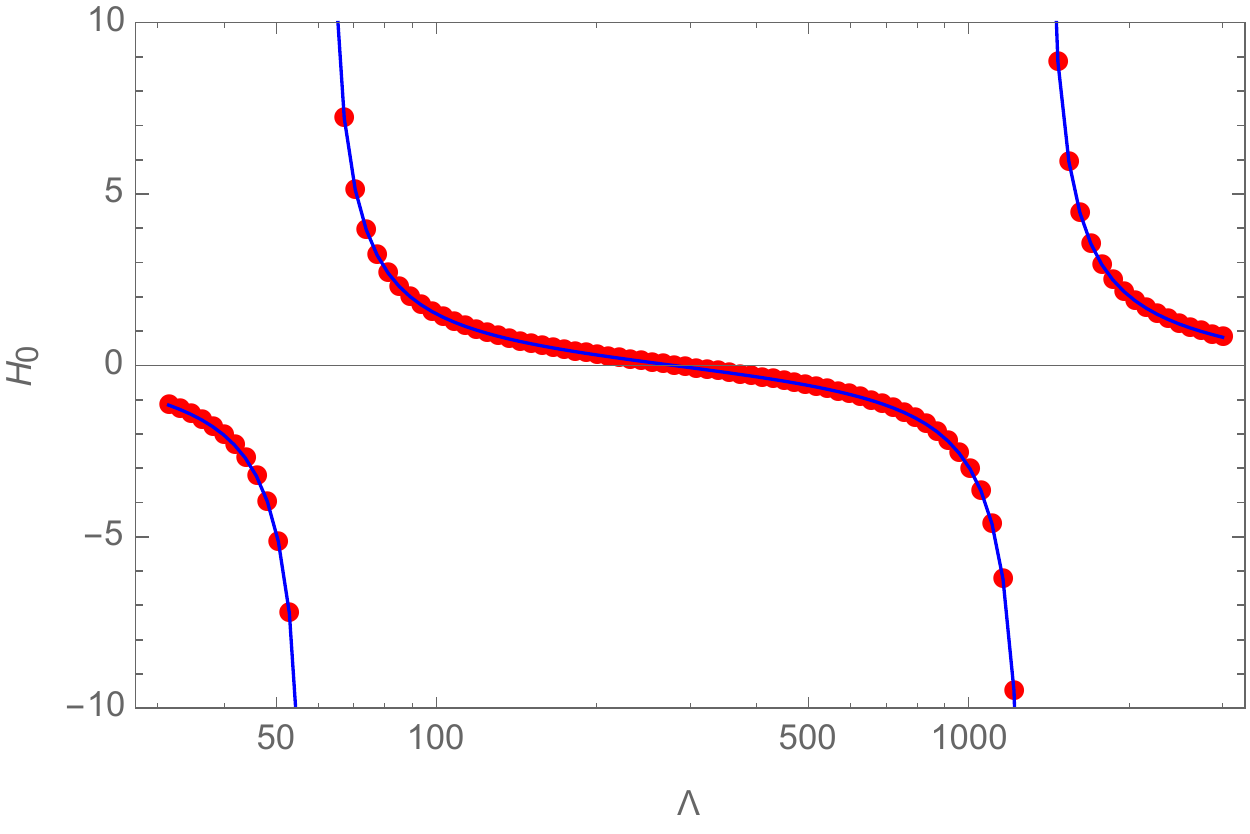}
\par
\end{centering}
\caption{\label{fig:Running}
  Running of $H_{0}(\Lambda)$ for $B_3=10$. Note that we take units
with $m=a=1$.}
\end{figure}

What would happen, if one would choose another cutoff? Then, in order to obtain
the same value of $B_3$, one would have to readjust $H_0(\Lambda)$. The outcome
is shown in Fig.~\ref{fig:Running}. It is clearly 
seen that the $\Lambda$-dependence of $H_0(\Lambda)$ exhibits the
characteristic log-periodic behavior~\cite{Bedaque:1998km,Bedaque:1998kg}. 
The essential point here is that fixing of {\em one} coupling constant
guarantees the cutoff-independence of the whole tower of the Efimov 
states~\cite{Efimov}, as well as the particle-dimer scattering amplitude 
above threshold (up to the corrections of order of $1/\Lambda$).
In the context of the problem we are considering, the study of the
$\Lambda$-dependence is intertwined with the study of the large-$L$
behavior -- recall that the dimension of the matrix in the quantization 
condition given in Eq.~(\ref{eq:master}) is $N^3\times N^3$, where $N\simeq L\Lambda/(2\pi)$. Consequently, in order to study the {\em scattering states} in the limit $L\to\infty$ while keeping $N$ finite, one has to consider the 
small values of $\Lambda$ as well. Note here that the cutoff effects will
not necessarily become large, since the energy of scattering states also 
tends to zero in this limit. 
 
Let us now turn to the spectrum of the model -- first, in the infinite volume:
\begin{enumerate}[leftmargin=0.5cm]
\item
  In the two-body subsystem, there exists a bound dimer. The binding energy is
\eq
B_2=\frac{1}{ma^2}=1\, .
\en
The typical momentum of a dimer is $1/a=1$ and its
typical size is $a=1$.
\item
We have adjusted the parameters such that there exists a three-particle bound
state with the binding energy $B_3=10$. The corresponding bound-state momentum
is $\kappa=\sqrt{mB_3}\simeq 3.162$ and the typical size is the inverse of this
value. Since the characteristic size of the three-particle bound state is
significantly smaller than that of the dimer, this state, to a certain approximation,
 can be considered as a bound state of three-particles (without clustering 
into a particle and a dimer).

\item
One finds another bound state at $B_3=1.016$, corresponding to the bound-state
momentum $\kappa=1.008$. We have now $\kappa^2-a^{-2}\ll \kappa^2$.
Therefore, this
state, to a good approximation, can be considered as a loosely 
bound state of a particle and a tightly bound dimer (see, e.g., 
Ref.~\cite{pang1}).

Note also that the ratio of two energies $10/1.016\simeq 10$ strongly deviates
from the value $515.03$, which is predicted by universality in the limit of
infinitely large scattering length, $a\to\infty$. The typical size of the state with smaller binding energy, however, is much larger than the scattering length $a$.
\end{enumerate}

Thus, in the infinite volume, we have a bound dimer and two three-body 
bound states. In a finite volume, we expect:
\begin{enumerate}[leftmargin=0.5cm]
\item
Two bound states, with the energies which are equal to those in the continuum up
to exponentially suppressed corrections.
\item
A tower of the particle-dimer scattering states in the CM frame, with the 
particle and dimer having quantized back-to-back momenta. In the limit 
$L\to\infty$, all these states tend to the threshold $E=-1$ (the sum of the
particle and dimer masses minus $3m$).
\item
A tower of three-particle scattering states with zero total momentum. In the
limit $L\to\infty$, all these states should approach the threshold $E=0$.
\end{enumerate}
In the actual calculations all energies will be slightly displaced from the
free values, due to the interactions among the particles.

\subsection{The entire energy spectrum}

In the following, we will examine whether our expectation is indeed realized in the 
  toy model defined in Eqs.~(\ref{eq:tauL},\ref{eq:ERE},\ref{eq:Z}).
  To do so, we project the quantization condition~\eqref{eq:qcond2}
  to the irrep $A_1^+$, see Eq.~\eqref{eq:qc-gnl}.
The projection of the kernel onto this irrep, see Eq.~\eqref{eq:longproj}, gives
\begin{align}
Z^{A_{1}^{+}}(r,s)= & \sum_{g\in {\cal G}}Z(g\mathbf{p}_{0}(r),\mathbf{k}_{0}(s)),
\end{align}
which will be used in the three-body quantization condition,
  Eq.~(\ref{eq:qc-gnl}).
The indices $\sigma,\rho$ are dropped since the irrep $A_{1}^{+}$
is one-dimensional.
\begin{figure}[h]
\begin{centering}
\includegraphics[width=1\linewidth]{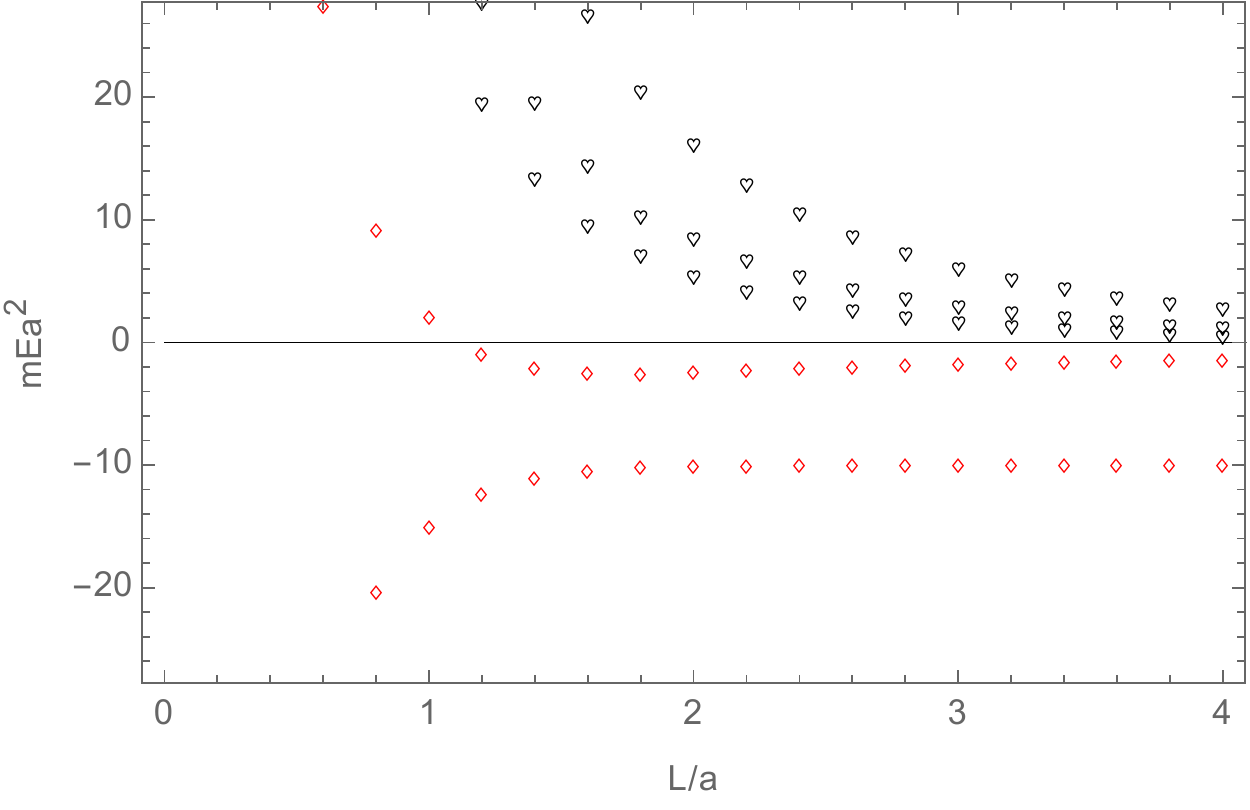}
\par\end{centering}
\caption{
\label{fig:entire}
$L$-dependence of the energy levels of the toy model projected to $A_1^+$ irrep, determined from the solutions of Eq.~\eqref{eq:qc-gnl} both below and above the threshold. Red diamonds and black hearts denote bound states and scattering states, respectively.
}
\end{figure}
The entire spectrum (lowest-lying levels), obtained from the solution
of the quantization condition is shown in Fig.~\ref{fig:entire} for different 
values of $L$.

\subsection{Bound states}

The two lowest levels in Fig.~\ref{fig:entire} tend to $E=-10$ and 
$E=-1.016$, respectively. As discussed above, the lowest energy level must look more like
a bound state of three particles, so one expects that 
its $L$-dependence is described
by the formula derived in Ref.~\cite{Meissner:2014dea}, see also 
Refs.~\cite{Hansen:2016ync,pang1}:
\eq\label{eq:three}
E_L-E_\infty=\frac{C}{L ^{3/2}}\,\exp\biggl(-\frac{2}{\sqrt{3}}\,\kappa L\biggr)\, ,\quad C<0\, .
\en
 From this formula it is seen that the energy level should approach
the infinite-volume limit from below. The second bound state is predominately
a bound state of a particle and a tightly bound dimer, so its $L$-dependence
is governed by the two-body L\"uscher formula~\cite{luescher-1}, see
also Refs.~\cite{pang1,Konig:2017krd}. The shift is again negative:
\eq\label{eq:particle-dimer}
E_L-E_\infty=\frac{C'}{L}\,\exp\biggl(-\frac{2}{\sqrt{3}}\,\sqrt{\kappa^2-a^{-2}} L\biggr)\,,~ C'<0~~~\, .
\en

\begin{figure*}
\begin{centering}
\includegraphics[width=0.49\linewidth]{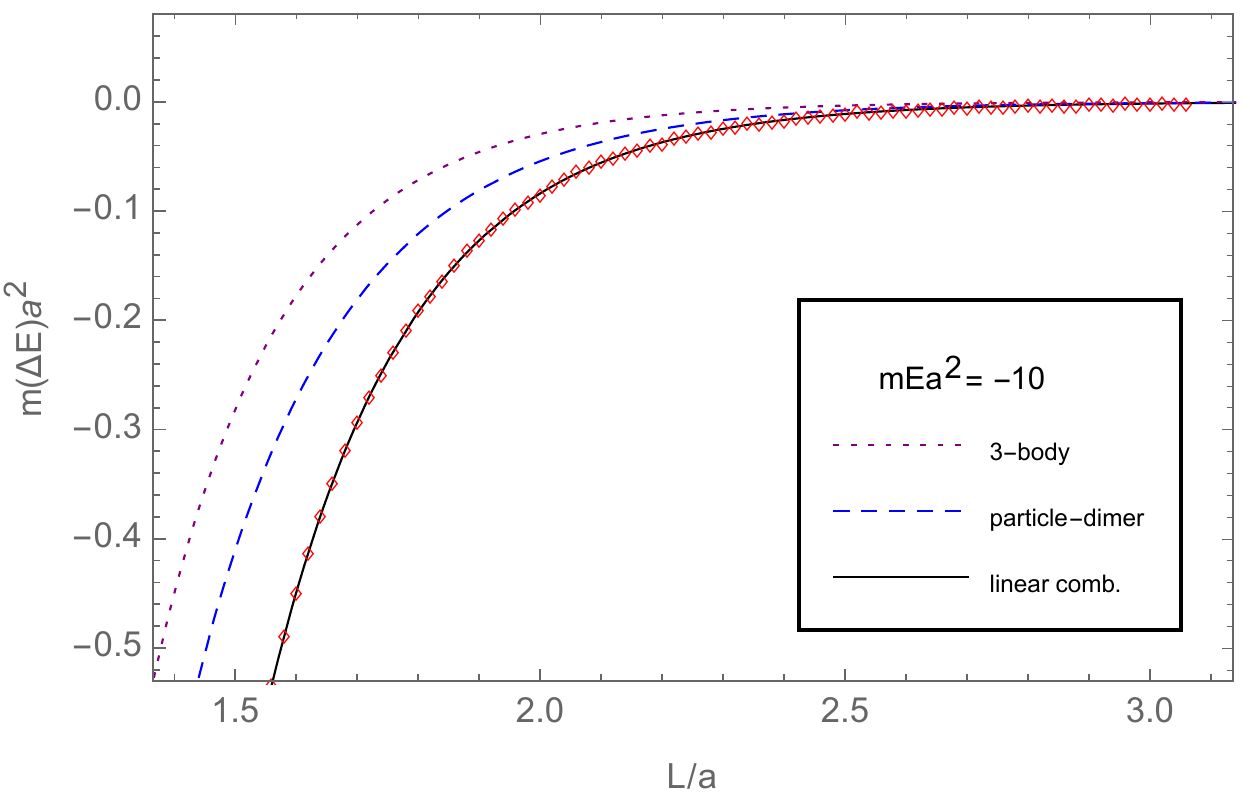}
\includegraphics[width=0.49\linewidth]{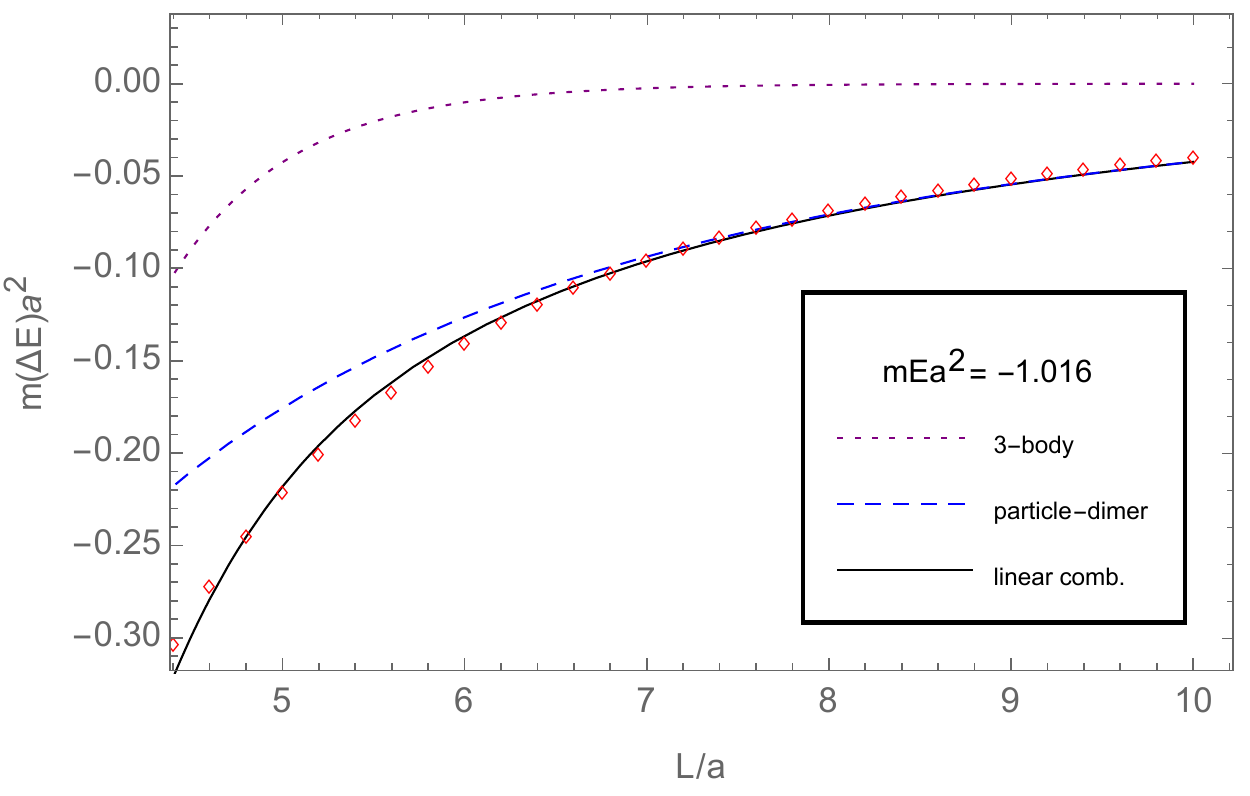}
\end{centering}
\caption{
\label{fig:bound}
The two lowest energy levels from Fig.~\ref{fig:entire}, corresponding to the
bound states (red diamonds). We show the results of the fit by using the linear combination of Eqs.~(\ref{eq:three}) and (\ref{eq:particle-dimer}), as well as the individual contributions.}
\end{figure*}

\begin{figure*}
\begin{centering}
\includegraphics[width=0.49\linewidth]{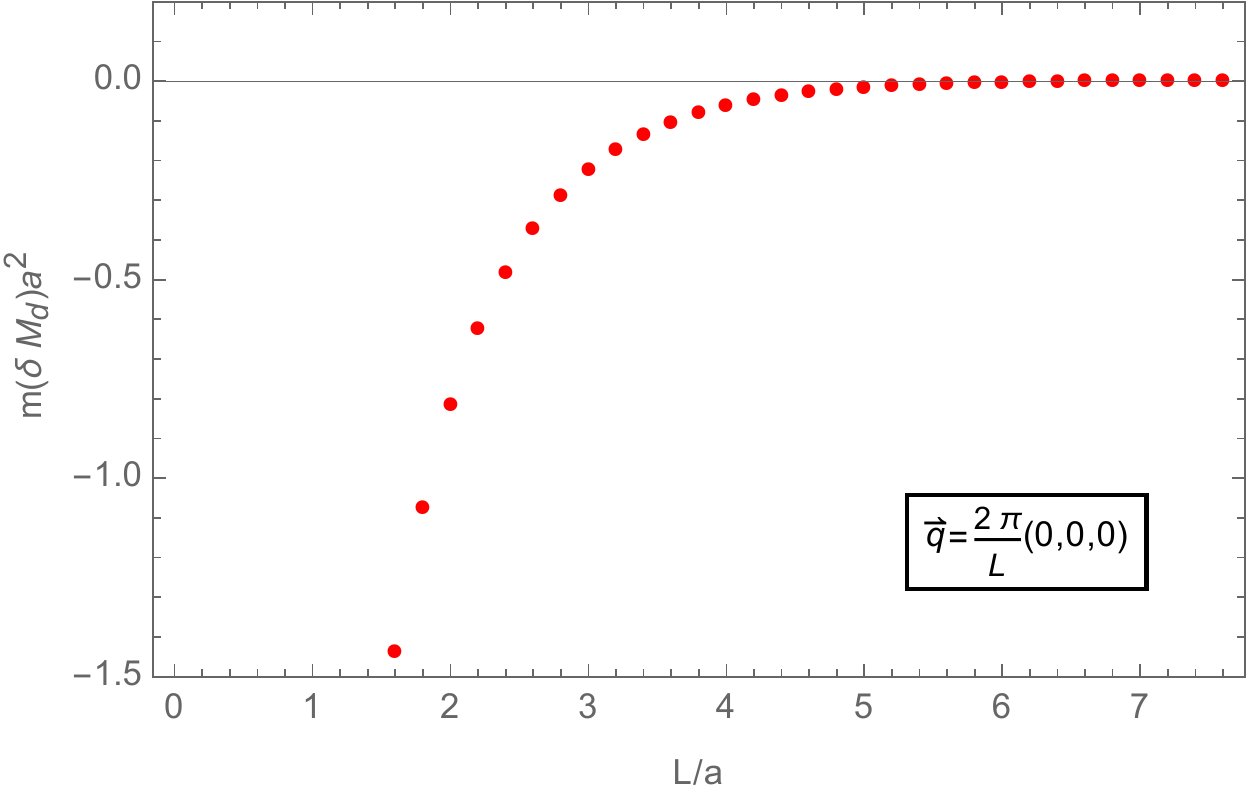}
\includegraphics[width=0.49\linewidth]{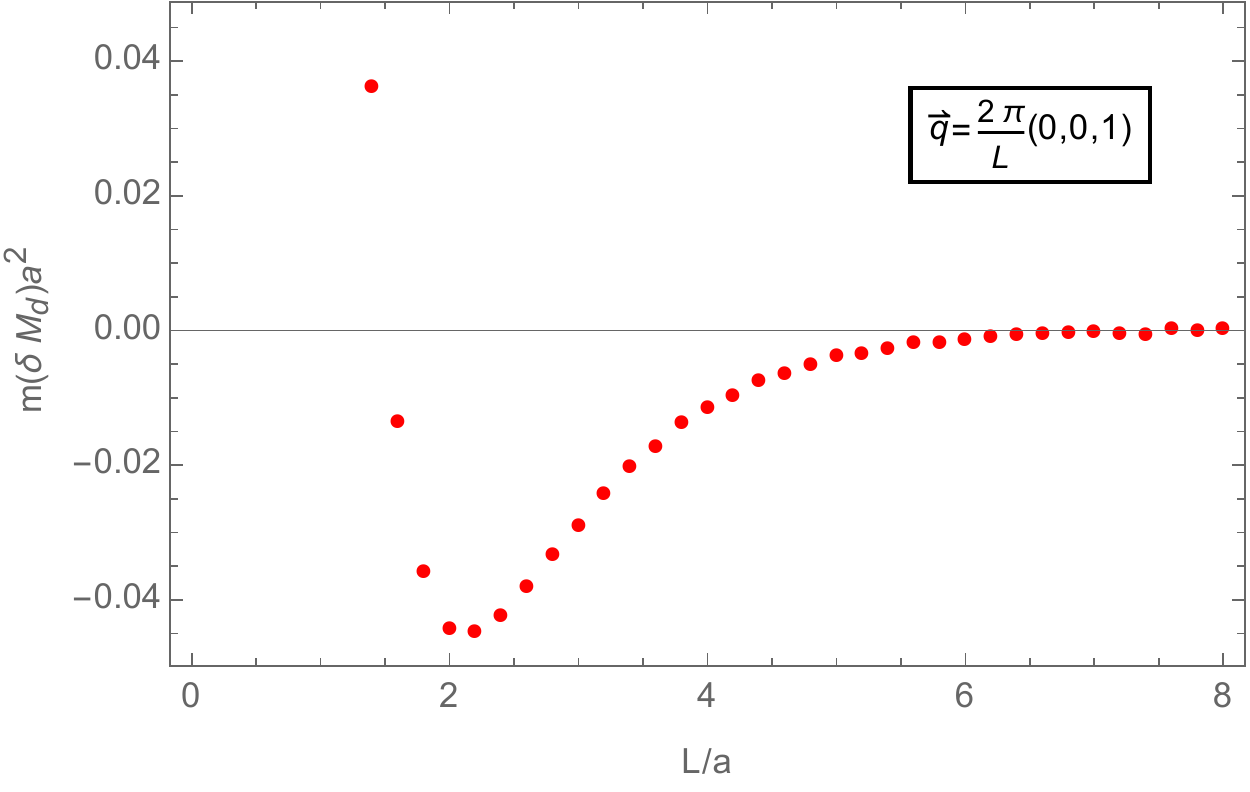}
\end{centering}
\caption{
\label{fig:dimer-energy}
Finite-volume corrections $\delta M_d=M_d(L)-M_d(\infty)$ to the dimer mass in rest frame (left panel) as well as in the
moving frame (right panel).}
\end{figure*}

The volume dependence of the bound-state energy levels is shown 
in Fig.~\ref{fig:bound}. We also show the fit to these energy levels 
by using the linear combination of
Eqs.~(\ref{eq:three}) and (\ref{eq:particle-dimer}), 
treating $C$ and $C'$ as free parameters, see Ref.~\cite{pang1}. 
The fitting range for $L$ is chosen to be 
$[1.5,3.0]$ and  $[4.6,10]$
in the left panel and right panel, respectively (it is necessary to choose different ranges
 because $\sqrt{\kappa^2-a^{-2}}\simeq 0.15\ll \kappa$).
In addition, we display two different
 contributions to the final fit separately. Here one sees that
the deeply bound state is well described by a mixture of the three-body 
bound state and a particle-dimer bound state with roughly equal weights
in magnitude, whereas the shallow one is predominately a particle-dimer
bound state. Moreover, we have checked that this picture stays robust,
when one increases the lower range of the fit interval (moving this range
to lower values of $L$ is not possible, because the suppressed contributions start to be sizable and the fit results are
not stable anymore). 
  Such a different behavior of two states can be understood as follows. First,
  we have seen before that the shallow bound state can be considered, to a good
  approximation, as a loosely
  bound state of a particle and a tightly bound dimer, since
  $\kappa ^2-a^{-2}\ll\kappa ^2$. At the same time, the deeply bound state can be
  considered as a predominately three-particle bound state based on the
  inequality $1/\kappa\ll a$. 
  Since the inequality is fulfilled better for the shallow state than for the deep one,
  the particle-dimer component is stronger in the shallow state than the three-particle
  component in the deep one. So, the obtained results
 perfectly agree with our expectations.

\begin{figure*}[t]
\begin{centering}
\includegraphics[width=0.48\linewidth]{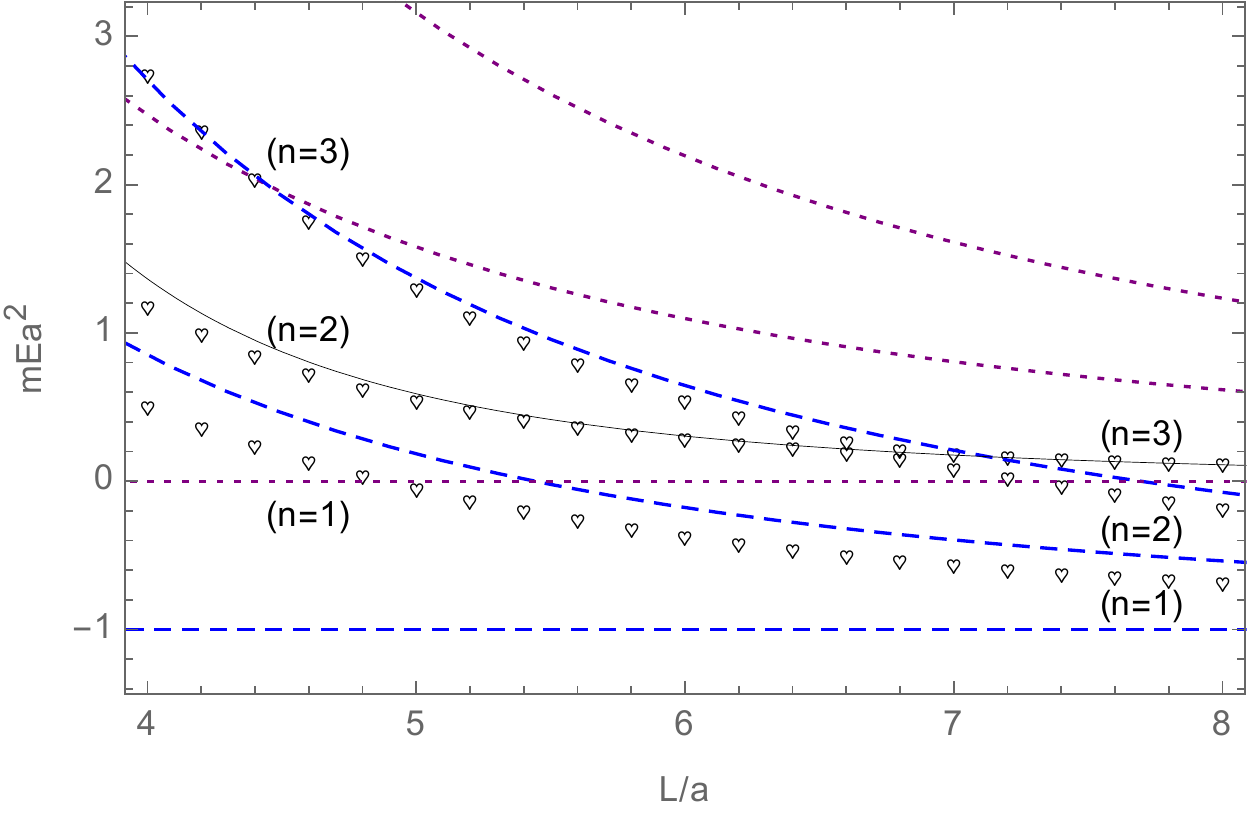}
\includegraphics[width=0.49\linewidth]{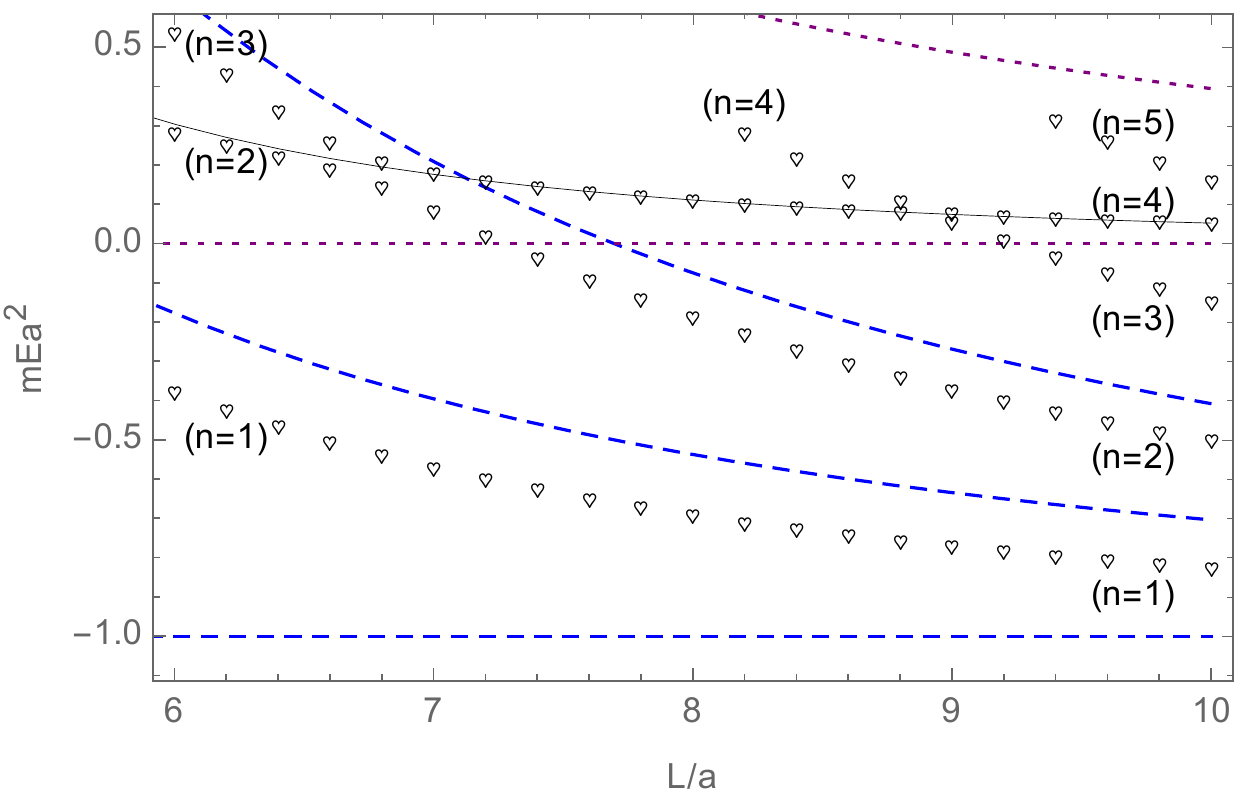}
\end{centering}
\caption{
\label{fig:groundstate}
The three lowest-lying scattering states (black hearts) above threshold for different plot ranges. Each energy level is labeled by ``(n=1,2,...)''. The result obtained using Eq.~(\ref{eq:Beane-Sharpe}) is given by the black solid curve.
The blue dashed curve shows the free particle-dimer states with back-to-back 
momenta $(0,0,0)$, $(0,0,1)$ and $(1,1,0)$, whereas the purple dotted lines
denote the free three-particle states (the lowest level at $E=0$ corresponds to the threshold, where all three particles are at rest).
}
\end{figure*}

\subsection{Scattering states}

Let us start from the mass of an isolated dimer. In Fig.~\ref{fig:dimer-energy}, 
left panel, we show the finite-volume corrections to the dimer mass in the rest frame, 
whereas in the right panel of the same
figure, we show this quantity in the moving frame, when the CM of a dimer
moves with a total momentum ${\bf k}=\frac{2\pi}{L}\,(0,0,1)$. The dimer mass is
defined as
\eq
E_2(L)=M_d(L)+\frac{{\bf k}^2}{4m}\, ,
\en
where $E_2(L)$ is the energy level in the two-particle system
(the solution of the  L\"uscher equation) and the
rest mass of the dimer constituents is not included in $M_d(L)$.
As seen from these figures,
the correction is the largest in the rest frame. This directly follows from the L\"uscher
equation, since the finite volume corrections vanish exponentially at large $L$.
The argument of the leading exponential is proportional to  
$k_dL$ with $k_d=\sqrt{\frac{1}{4}\,{\bf k}^2+mB_2}$
reaching its minimal value in the rest frame.
Furthermore, since the energy of the particle-dimer state is given by the particle
mass plus dimer mass plus the interaction energy between the particle and a dimer, one
may expect that the corrections due to the $L$-dependence of the dimer mass
are largest, when the dimer is in the rest frame.

Now let us return to Fig.~\ref{fig:entire} and try to interpret the states above threshold
as the particle-dimer and three-particle scattering states. At large $L$, the energies
of these states tend to $E=-1$ and $E=0$, respectively. In order to identify the levels at the 
intermediate values of $L$, let us use the expression for the volume-dependent
shift of the three-particle ground-state energy level, obtained, e.g., in Refs.~\cite{Beane:2007qr,Sharpe:2017jej}. Up to and including order $L^{-5}$, this expression 
contains only a single parameter $a$ and reads as
\eq\label{eq:Beane-Sharpe}
E=\frac{12\pi a}{L^3}-\frac{12 a^2}{L^4}\,{\cal I}
+\frac{12 a^3}{\pi L^5}({\cal I}^2+{\cal J})+O(L^{-6})\,,~~~
\en
where ${\cal I}\simeq -8.914$ and ${\cal J}\simeq 16.532$ are numerical constants.
We now plot again the three lowest eigenvalues above threshold and confront them
with the curve obtained from Eq.~(\ref{eq:Beane-Sharpe}), see Fig.~\ref{fig:groundstate}.

%

It is seen that, below $L\simeq 6.5$, the second level closely follows the prediction of the formula, so that it can be identified with the (shifted) ground state. After $L\simeq 6.5$ an avoided level crossing occurs, and it becomes clear that the third level has 
to be interpreted as the shifted ground state. The other two levels continue to move 
towards $E=-1$ and from now on 
should be interpreted as the particle-dimer scattering states. 
This is seen even better in the right panel of Fig.~\ref{fig:groundstate}, which covers a larger 
interval in $L$. We observe more than one avoided level crossing in this figure.

Furthermore, as seen from these figures, the spectrum below the three-particle 
threshold closely follows the free particle-dimer energy levels (a small displacement is 
caused by interactions). However, at first glance, it 
seems that the counterpart of the ground-state level with the vanishing back-to-back 
momentum is missing. 

\begin{figure}[t]
\begin{centering}
\includegraphics[width=1\linewidth]{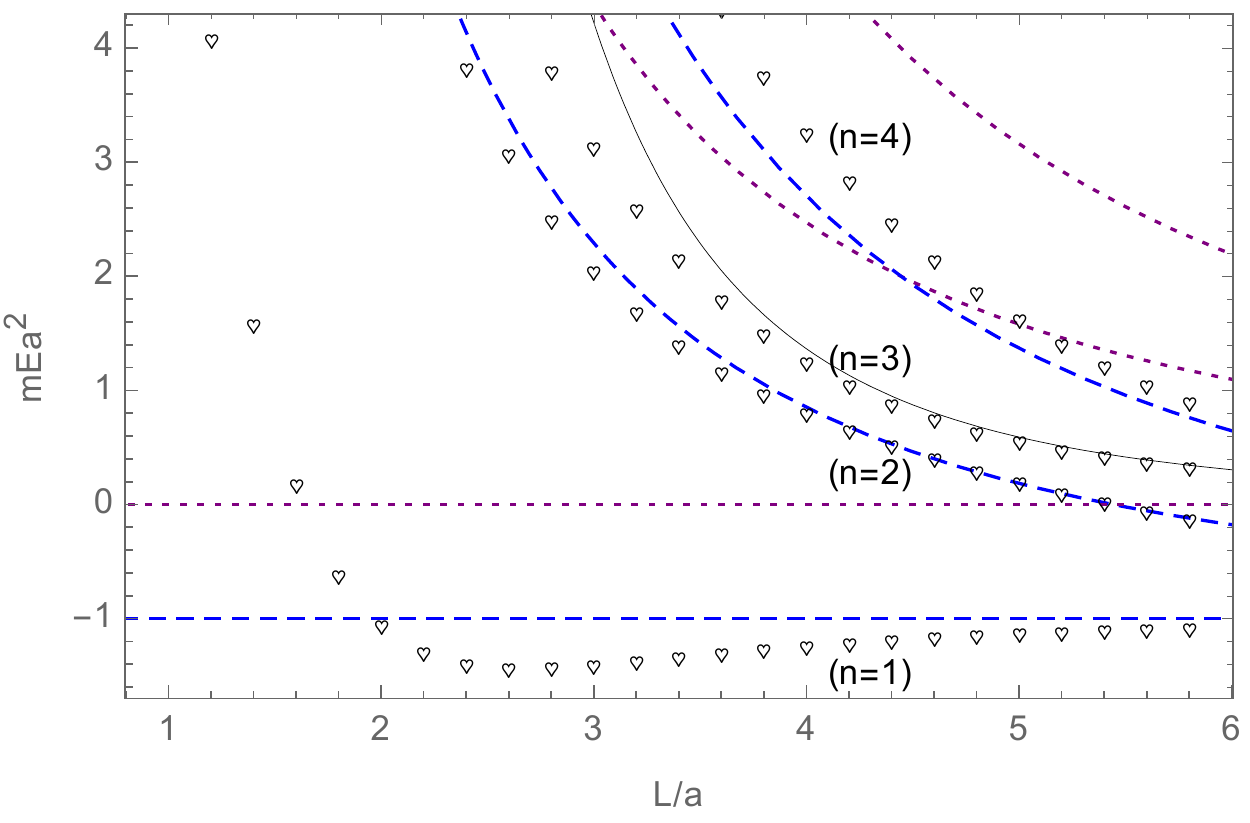}
\par\end{centering}
\caption{\label{fig:without}
The same as in Fig.~\ref{fig:groundstate}, but
with a different value of the parameter $H_0(\Lambda)$, for which the shallow
three-body
bound state does not exist. The level that lies in the vicinity of
 the free particle-dimer threshold is a {\em scattering state} now.}
\end{figure}

In order to explain such a seemingly strange behavior, note that in the model
we have a very shallow particle-dimer ground state with the energy $E=-1.016$, which 
``pushes'' the next level up, close to the next free level with back-to-back momentum $(0,0,1)$. This can be seen in the following manner. 
One could choose the parameters of the model, so that the shallow bound state disappears. Choosing, for example, $H_0(\Lambda)=-1.353$ and the cutoff $\Lambda=20$, we get the energy of the deep bound state equal to $E=-5$, whereas the shallow bound state does not appear in the infinite-volume spectrum any more. The finite-volume energy spectrum for this case is shown 
in Fig.~\ref{fig:without}, where the level in the vicinity of the particle-dimer threshold is now a {\em scattering state}. Changing now all parameters of the model continuously, so that the lowest state just becomes
the bound state again,  will not change the position of the other levels very much, so in this case the level in the vicinity of the back-to-back momentum  $(0,0,1)$ indeed corresponds to the lowest scattering state.

Albeit the intuitive interpretation of the energy levels, which was given above, is very transparent, we would like to stress that it is good for illustrative purposes only. Strictly speaking, in a finite volume one has only a spectrum that is determined by a full Hamiltonian of the system -- no further labeling of
the states can be justified rigorously.

\section{Conclusion}

The main results of our work can be summarized as follows:

\begin{itemize}[leftmargin=0.5cm]
\item[(i)]
We have performed the projection of the three-particle quantization condition
onto the different irreps of the octahedral group.
As a result, the quantization condition
partially diagonalizes -- the different irreps do not talk to each other. Two alternative methods
have been proposed. In both methods, the kernel that enters the quantization condition, instead
of the spherical functions, is expanded in the basis vectors of various irreps of the octahedral
group. The corresponding quantization conditions,
  given by Eqs.~\eqref{eq:qc-gnl} and \eqref{eq:qcond2}, are essentially equivalent.
\item[(ii)]
  Using this method, the finite volume spectrum in the $A_1^+$ irrep was calculated, using a toy model corresponding to the leading order of the effective field theory for short range-interactions~\cite{Bedaque:1998kg,Bedaque:1998km}. This model has interactions in the $S$-wave only. The results are very instructive. One directly sees that different bound states of the same model may have different nature (a three-particle bound state, a particle-dimer bound state, or something in between). Moreover, the scattering states cannot be uniquely interpreted as particle-dimer scattering states (in the infinite volume, these
states appear in the elastic and rearrangement channels) and three-particle states (appearing in breakup reactions). At certain values of $L$, an ``avoided level crossing''
takes place when the different energy levels change their roles.
\end{itemize}

In the future, one may consider generalizations of this method in several directions. In particular, one may include interactions in higher partial waves, both in the two-body and three-body (particle-dimer) sector.  One may consider moving frames. Finally, one may consider particles with   spin, with the case of three nucleons being most interesting. 
Work in these directions is underway and will be discussed in future publications.

\subsection*{Acknowledgments}

The authors thank U.-G. Mei{\ss}ner and E. Epelbaum for useful discussions. MD, MM, and AR acknowledge fruitful discussions with I. Aitchison, R. Brice\~no, Z. Davoudi, M. Hansen, A. Szczepaniak and S. Sharpe during the INT Workshop ``Multi-Hadron Systems from Lattice QCD.''  We acknowledge the support from the CRC 110
``Symmetries and the Emergence of Structure in QCD''
 (DFG grant no. TRR~110 and NSFC grant no. 11621131001)
 and the CRC 1245 ``Nuclei: From Fundamental Interactions to Structure
 and Stars'' (DFG grant no. SFB 1245) as well as support from
 the BMBF under contract no. 05P15RDFN1.
This work was supported by the National Science Foundation under Grants No. 
PHY-1452055 and PHY-1415459 and by the U.S. Department of Energy, Office of Science, Office of Nuclear Physics
under award number DE-SC001658 and under contract number DE-AC05-06OR23177. 
MM is thankful to the German Research Foundation (DFG) for the financial support, under the
fellowship MA 7156/1-1, as well as to the George Washington
University for the hospitality and inspiring environment.
 This research was also supported in part by Volkswagenstiftung under contract no. 93562
 and by Shota Rustaveli National Science Foundation (SRNSF), grant no. DI-2016-26. 
MD, HWH, JYP, and AR thank the German Research Foundation (DFG) for
 support during the 2018 Hirschegg workshop ``Multiparticle resonances in hadrons, nuclei, and ultracold gases'' where this work was finalized.

\appendix
\begin{widetext}

\newpage
\section{Matrices of the irreducible representations}
\label{app:group}

\newcommand{\XA}{$\begin{pmatrix} 1 & 0\cr 0 & 1\end{pmatrix}$}
\newcommand{\XB}{$\begin{pmatrix} -\frac{1}{2} & \frac{\sqrt{3}}{2}\cr -\frac{\sqrt{3}}{2} & -\frac{1}{2}\end{pmatrix}$}
\newcommand{\XC}{$\begin{pmatrix} -\frac{1}{2} & -\frac{\sqrt{3}}{2}\cr \frac{\sqrt{3}}{2} & -\frac{1}{2}\end{pmatrix}$}
\newcommand{\XD}{$\begin{pmatrix} -\frac{1}{2} & -\frac{\sqrt{3}}{2}\cr -\frac{\sqrt{3}}{2} & \frac{1}{2}\end{pmatrix}$}
\newcommand{\XE}{$\begin{pmatrix} 1 & 0\cr 0 & -1\end{pmatrix}$}
\newcommand{\XF}{$\begin{pmatrix} -\frac{1}{2} & \frac{\sqrt{3}}{2}\cr \frac{\sqrt{3}}{2} & \frac{1}{2}\end{pmatrix}$}

In the following table, we quote the matrix representations of the five irreps $A_1,A_2,E,T_1,T_2$.
The numbering of the group elements $a=1,\ldots,24$ corresponds to 
table~A.1 of Ref.~\cite{Bernard:2008ax}.\\

\renewcommand{\arraystretch}{1.6}
\begin{table}[h]
\resizebox{0.44\linewidth}{!}{
\begin{tabular}[t]{|c| c| c| c| c| c| c|}
\hline
$a$ & CC& $A_1$ & $A_2$ & $E$ & $T_1$ & $T_2$ \\
\hline
1 & $I$ & $1$ & $1$ & \XA & 
$\begin{pmatrix} 1 & 0 & 0 \cr 0 & 1 & 0 \cr 0 & 0 & 1\end{pmatrix}$ & 
$\begin{pmatrix} 1 & 0 & 0 \cr 0 & 1 & 0 \cr 0 & 0 & 1\end{pmatrix}$ \\
\hline
2 & $8C_3$ & $1$ & $1$ & \XB &
$\begin{pmatrix} 0 & 1 & 0 \cr 0 & 0 & 1 \cr 1 & 0 & 0\end{pmatrix}$ & 
$\begin{pmatrix} 0 & 1 & 0 \cr 0 & 0 & 1 \cr 1 & 0 & 0\end{pmatrix}$ \\
3 & & $1$ & $1$ & \XC &
$\begin{pmatrix} 0 & 0 & 1 \cr 1 & 0 & 0 \cr 0 & 1 & 0\end{pmatrix}$ & 
$\begin{pmatrix} 0 & 0 & 1 \cr 1 & 0 & 0 \cr 0 & 1 & 0\end{pmatrix}$ \\
4 & & $1$ & $1$ & \XC &
$\begin{pmatrix} 0 & 0 & -1 \cr -1 & 0 & 0 \cr 0 & 1 & 0\end{pmatrix}$ & 
$\begin{pmatrix} 0 & 0 & -1 \cr -1 & 0 & 0 \cr 0 & 1 & 0\end{pmatrix}$ \\ 
5 & & $1$ & $1$ & \XB &
$\begin{pmatrix} 0 & -1 & 0 \cr 0 & 0 & 1 \cr -1 & 0 & 0\end{pmatrix}$ & 
$\begin{pmatrix} 0 & -1 & 0 \cr 0 & 0 & 1 \cr -1 & 0 & 0\end{pmatrix}$ \\
6 & & $1$ & $1$ & \XB &
$\begin{pmatrix} 0 & 1 & 0 \cr 0 & 0 & -1 \cr -1 & 0 & 0\end{pmatrix}$ & 
$\begin{pmatrix} 0 & 1 & 0 \cr 0 & 0 & -1 \cr -1 & 0 & 0\end{pmatrix}$ \\
7 & & $1$ & $1$ & \XC &
$\begin{pmatrix} 0 & 0 & -1 \cr 1 & 0 & 0 \cr 0 & -1 & 0\end{pmatrix}$ & 
$\begin{pmatrix} 0 & 0 & -1 \cr 1 & 0 & 0 \cr 0 & -1 & 0\end{pmatrix}$ \\
8 & & $1$ & $1$ & \XC &
$\begin{pmatrix} 0 & 0 & 1 \cr -1 & 0 & 0 \cr 0 & -1 & 0\end{pmatrix}$ & 
$\begin{pmatrix} 0 & 0 & 1 \cr -1 & 0 & 0 \cr 0 & -1 & 0\end{pmatrix}$ \\
9 & & $1$ & $1$ & \XB &
$\begin{pmatrix} 0 & -1 & 0 \cr 0 & 0 & -1 \cr 1 & 0 & 0\end{pmatrix}$ & 
$\begin{pmatrix} 0 & -1 & 0 \cr 0 & 0 & -1 \cr 1 & 0 & 0\end{pmatrix}$ \\
\hline
10 & $6C_4$ & $1$ & $-1$ & \XD &
$\begin{pmatrix} 1 & 0 & 0 \cr 0 & 0 & 1 \cr 0 & -1 & 0\end{pmatrix}$ & 
$\begin{pmatrix} -1 & 0 & 0 \cr 0 & 0 & -1 \cr 0 & 1 & 0\end{pmatrix}$ \\ 
11 & & $1$ & $-1$ & \XD &
$\begin{pmatrix} 1 & 0 & 0 \cr 0 & 0 & -1 \cr 0 & 1 & 0\end{pmatrix}$ & 
$\begin{pmatrix} -1 & 0 & 0 \cr 0 & 0 & 1 \cr 0 & -1 & 0\end{pmatrix}$ \\
12 & & $1$ & $-1$ & \XF &
$\begin{pmatrix} 0 & 0 & -1 \cr 0 & 1 & 0 \cr 1 & 0 & 0\end{pmatrix}$ & 
$\begin{pmatrix} 0 & 0 & 1 \cr 0 & -1 & 0 \cr -1 & 0 & 0\end{pmatrix}$ \\
\hline
\end{tabular}
}
~~~
\resizebox{0.44\linewidth}{!}{
\begin{tabular}[t]{|c| c| c| c| c| c| c|}
\hline
$a$ & CC& $A_1$ & $A_2$ & $E$ & $T_1$ & $T_2$ \\
\hline 
13 & & $1$ & $-1$ & \XF &
$\begin{pmatrix} 0 & 0 & 1 \cr 0 & 1 & 0 \cr -1 & 0 & 0\end{pmatrix}$ & 
$\begin{pmatrix} 0 & 0 & -1 \cr 0 & -1 & 0 \cr 1 & 0 & 0\end{pmatrix}$ \\
14 & & $1$ & $-1$ & \XE &
$\begin{pmatrix} 0 & 1 & 0 \cr -1 & 0 & 0 \cr 0 & 0 & 1\end{pmatrix}$ & 
$\begin{pmatrix} 0 & -1 & 0 \cr 1 & 0 & 0 \cr 0 & 0 & -1\end{pmatrix}$ \\
15 & & $1$ & $-1$ & \XE &
$\begin{pmatrix} 0 & -1 & 0 \cr 1 & 0 & 0 \cr 0 & 0 & 1\end{pmatrix}$ & 
$\begin{pmatrix} 0 & 1 & 0 \cr -1 & 0 & 0 \cr 0 & 0 & -1\end{pmatrix}$ \\
\hline
16 & $6C_2'$ & $1$ & $-1$ & \XD &
$\begin{pmatrix} -1 & 0 & 0 \cr 0 & 0 & 1 \cr 0 & 1 & 0\end{pmatrix}$ & 
$\begin{pmatrix} 1 & 0 & 0 \cr 0 & 0 & -1 \cr 0 & -1 & 0\end{pmatrix}$ \\
17 & & $1$ & $-1$ & \XD &
$\begin{pmatrix} -1 & 0 & 0 \cr 0 & 0 & -1 \cr 0 & -1 & 0\end{pmatrix}$ & 
$\begin{pmatrix} 1 & 0 & 0 \cr 0 & 0 & 1 \cr 0 & 1 & 0\end{pmatrix}$ \\ 
18 & & $1$ & $-1$ & \XE &
$\begin{pmatrix} 0 & 1 & 0 \cr 1 & 0 & 0 \cr 0 & 0 & -1\end{pmatrix}$ & 
$\begin{pmatrix} 0 & -1 & 0 \cr -1 & 0 & 0 \cr 0 & 0 & 1\end{pmatrix}$ \\
19 & & $1$ & $-1$ & \XE &
$\begin{pmatrix} 0 & -1 & 0 \cr -1 & 0 & 0 \cr 0 & 0 & -1\end{pmatrix}$ & 
$\begin{pmatrix} 0 & 1 & 0 \cr 1 & 0 & 0 \cr 0 & 0 & 1\end{pmatrix}$ \\ 
20 & & $1$ & $-1$ & \XF &
$\begin{pmatrix} 0 & 0 & 1 \cr 0 & -1 & 0 \cr 1 & 0 & 0\end{pmatrix}$ & 
$\begin{pmatrix} 0 & 0 & -1 \cr 0 & 1 & 0 \cr -1 & 0 & 0\end{pmatrix}$ \\
21 & & $1$ & $-1$ & \XF &
$\begin{pmatrix} 0 & 0 & -1 \cr 0 & -1 & 0 \cr -1 & 0 & 0\end{pmatrix}$ & 
$\begin{pmatrix} 0 & 0 & 1 \cr 0 & 1 & 0 \cr 1 & 0 & 0\end{pmatrix}$ \\ 
\hline
22 & $3C_2$ & $1$ & $1$ & \XA &
$\begin{pmatrix} 1 & 0 & 0 \cr 0 & -1 & 0 \cr 0 & 0 & -1\end{pmatrix}$ & 
$\begin{pmatrix} 1 & 0 & 0 \cr 0 & -1 & 0 \cr 0 & 0 & -1\end{pmatrix}$ \\ 
23 & & $1$ & $1$ & \XA &
$\begin{pmatrix} -1 & 0 & 0 \cr 0 & 1 & 0 \cr 0 & 0 & -1\end{pmatrix}$ & 
$\begin{pmatrix} -1 & 0 & 0 \cr 0 & 1 & 0 \cr 0 & 0 & -1\end{pmatrix}$ \\
24 & & $1$ & $1$ & \XA &
$\begin{pmatrix} -1 & 0 & 0 \cr 0 & -1 & 0 \cr 0 & 0 & 1\end{pmatrix}$ & 
$\begin{pmatrix} -1 & 0 & 0 \cr 0 & -1 & 0 \cr 0 & 0 & 1\end{pmatrix}$ \\
\hline
\end{tabular}
}
\caption{Matrix representations of
  the irreps of the octahedral group (pure rotations). ``CC`` denotes the conjugacy class, and $a=1,..,24$ index the group elements.}
\end{table}

\newpage

\section{Linear independent sets of cubic harmonics}
\label{app:cubic}

Maximal sets of linearly
  independent cubic harmonics for $\ell\le9$ as determined by the procedure described in Section.~\ref{sec:alternative} for the corresponding shell types.

\renewcommand{\arraystretch}{1.6}
\begin{table}[h]
\resizebox{\linewidth}{!}{%
\begin{tabular}{c  c  l}
\toprule
~Shell type~& ~$\vartheta$~ & Set\\
\colrule
$(000)$&$1$&$
X_ 0^{A_ 1^+ 11}$ \bigstrut[t] \\
\colrule
$(00b)$&$6$&$
X_ 0^{A_ 1^+ 11},X_ 2^{E^+ 11},X_ 2^{E^+ 12},X_ 1^{T_ 1^- 11},X_ 1^{T_ 1^- 12},X_ 1^{T_ 1^- 13}$ \bigstrut[t]\\
\colrule
$(0aa)$&$12$&$
X_ 0^{A_ 1^+ 11},X_ 2^{E^+ 11},X_ 2^{E^+ 12},X_ 1^{T_ 1^- 11},X_ 1^{T_ 1^- 12},X_ 1^{T_ 1^- 13},X_ 2^{T_ 2^+ 11},X_ 2^{T_ 2^+ 12},X_ 2^{T_ 2^+ 13},X_ 3^{T_ 2^- 11},X_ 3^{T_ 2^- 12},X_ 3^{T_ 2^- 13}$ \bigstrut[t]\\
\colrule
$(0bc)$&$24$&
$X_ 0^{A_ 1^+ 11},X_ 6^{A_ 2^+ 11},X_ 2^{E^+ 11},X_ 2^{E^+ 12},X_ 4^{E^+ 11},X_ 4^{E^+ 12},X_ 4^{T_ 1^+ 11},X_ 4^{T_ 1^+ 12},X_ 4^{T_ 1^+ 13},X_ 1^{T_ 1^- 11},X_ 1^{T_ 1^- 12},X_ 1^{T_ 1^- 13},$\bigstrut[t]\\
&&$X_ 3^{T_ 1^- 11},X_ 3^{T_ 1^- 12},X_ 3^{T_ 1^- 13},X_ 2^{T_ 2^+ 11},X_ 2^{T_ 2^+ 12},X_ 2^{T_ 2^+ 13},X_ 3^{T_ 2^- 11},X_ 3^{T_ 2^- 12},X_ 3^{T_ 2^- 13},X_ 5^{T_ 2^- 11},X_ 5^{T_ 2^- 12},X_ 5^{T_ 2^- 13}$ \\
\colrule
$(bbb)$&$8$&$
X_ 0^{A_ 1^+ 11},X_ 3^{A_ 2^- 11},X_ 1^{T_ 1^- 11},X_ 1^{T_ 1^- 12},X_ 1^{T_ 1^- 13},X_ 2^{T_ 2^+ 11},X_ 2^{T_ 2^+ 12},X_ 2^{T_ 2^+ 13}$ \bigstrut[t]\\
\colrule
$(bbc)$&24&$
X_ 0^{A_ 1^+ 11},X_ 3^{A_ 2^- 11},X_ 2^{E^+ 11},X_ 2^{E^+ 12},X_ 5^{E^- 11},X_ 5^{E^- 12},X_ 4^{T_ 1^+ 11},X_ 4^{T_ 1^+ 12},X_ 4^{T_ 1^+ 13},X_ 1^{T_ 1^- 11},X_ 1^{T_ 1^- 12},X_ 1^{T_ 1^- 13},$ \bigstrut[t]\\
&&$X_ 3^{T_ 1^- 11},X_ 3^{T_ 1^- 12},X_ 3^{T_ 1^- 13},X_ 2^{T_ 2^+ 11},X_ 2^{T_ 2^+ 12},X_ 2^{T_ 2^+ 13},X_ 4^{T_ 2^+ 11},X_ 4^{T_ 2^+ 12},X_ 4^{T_ 2^+ 13},X_ 3^{T_ 2^- 11},X_ 3^{T_ 2^- 12},X_ 3^{T_ 2^- 13}$\\
\colrule
$(bcd)$&$48$&
$X_ 0^{A_ 1^+ 11},X_ 9^{A_ 1^- 11},X_ 6^{A_ 2^+ 11},X_ 3^{A_ 2^- 11},X_ 2^{E^+ 11},X_ 2^{E^+ 12},X_ 4^{E^+ 11},X_ 4^{E^+ 12},X_ 5^{E^- 11},X_ 5^{E^- 12},X_ 7^{E^- 11},X_ 7^{E^- 12},$ \bigstrut[t]\\
&&$X_ 4^{T_ 1^+ 11},X_ 4^{T_ 1^+ 12},X_ 4^{T_ 1^+ 13},X_ 6^{T_ 1^+ 11},X_ 6^{T_ 1^+ 12},X_ 6^{T_ 1^+ 13},X_ 8^{T_ 1^+ 21},X_ 8^{T_ 1^+ 22},X_ 8^{T_ 1^+ 23},X_ 1^{T_ 1^- 11},X_ 1^{T_ 1^- 12},$\\
&&$X_ 1^{T_ 1^- 13},X_ 3^{T_ 1^- 11},X_ 3^{T_ 1^- 12},X_ 3^{T_ 1^- 13},X_ 5^{T_ 1^- 11},X_ 5^{T_ 1^- 12},X_ 5^{T_ 1^- 13},X_ 2^{T_ 2^+ 11},X_ 2^{T_ 2^+ 12},X_ 2^{T_ 2^+ 13},X_ 4^{T_ 2^+ 11},$\\
&&$X_ 4^{T_ 2^+ 12},X_ 4^{T_ 2^+ 13},X_ 6^{T_ 2^+ 11},X_ 6^{T_ 2^+ 12},X_ 6^{T_ 2^+ 13},X_ 3^{T_ 2^- 11},X_ 3^{T_ 2^- 12},X_ 3^{T_ 2^- 13},X_ 5^{T_ 2^- 11},X_ 5^{T_ 2^- 12},X_ 5^{T_ 2^- 13},$\\
&&$X_ 7^{T_ 2^- 21},X_ 7^{T_ 2^- 22},X_ 7^{T_ 2^- 23}$\\
\botrule
\end{tabular}
}
\caption{Full sets of cubic harmonics $X_\ell^{\Gamma\nu \alpha}$
  with $\ell\leq 9$, contributing to the basis vectors on a shell of a given type. Here $\vartheta$ denotes the multiplicity, i.e., the number of points (and therefore of basis vectors) on 
a given shell.}
\label{tab:shell}
\end{table}

\end{widetext}

\end{document}